\documentclass[12pt]{article}

\usepackage{booktabs} 
\usepackage{setspace} 
\usepackage{verbatim} 
\usepackage{url} 
\usepackage{graphicx} 
\usepackage{amsmath} 
\usepackage{amsfonts} 

\usepackage[ruled]{algorithm2e} 
\usepackage[justification=centering]{caption}

\newcommand{\stsp}{\ensuremath{\mathcal{S}}}
\newcommand{\acsp}{\ensuremath{\mathcal{A}}}
\newcommand{\mxstm}{\ensuremath{\mathit{STM}}}

\newcommand{\mxcard}[1]{\left|{#1}\right|}

\SetAlFnt{\small}
\SetAlCapFnt{\small}
\SetAlCapNameFnt{\small}
\SetAlCapHSkip{0pt}
\IncMargin{-\parindent}

\begin{document}

\begin{center}
\begin{spacing}{1.8}
{\Large\textbf{Runtime Adaptation in Wireless Sensor Nodes Using
Structured Learning}\footnote{This is a pre-publication version of a paper 
that has been
accepted for publication in the ACM Transactions on
Cyber-Physical Systems.
The official/final version of the paper
will be posted on the ACM Digital Library.}}\par
\end{spacing}
\vspace{.12in}
Adrian Sapio\textsuperscript{1}, Shuvra S.~Bhattacharyya\textsuperscript{1}, 
Marilyn Wolf\textsuperscript{2}\\
\vspace{.20in}
1.~University of Maryland\\[0.05cm]
College Park, Maryland, USA\\[0.05cm]
Email: \url{adrian.sapio@gmail.com}, \url{ssb@umd.edu}\\
\vspace{.14in}
2.~University of Nebraska\\[0.05cm]
Lincoln, Nebraska, USA\\[0.05cm]
Email: \url{mwolf@unl.edu}\\
\vspace{.12in}
\end{center}


\section*{Abstract} 
Markov Decision Processes (MDPs) provide important capabilities for
facilitating the dynamic adaptation and self-optimization of cyber physical 
systems at runtime. 
In recent years, this has primarily taken the form of Reinforcement Learning (RL)
techniques that eliminate some MDP components for the purpose of 
reducing computational requirements. In this work, we show that 
recent advancements in Compact MDP Models (CMMs) provide sufficient 
cause to question this trend when designing wireless sensor network
nodes. 
In this work, a novel CMM-based approach to designing self-aware wireless
sensor nodes is presented and compared to Q-Learning, a popular RL technique.
We show that a certain class of CPS nodes is not well served by RL methods, 
and contrast RL versus CMM methods in this context. 
Through both simulation and a prototype implementation, 
we demonstrate that CMM methods can provide significantly 
better runtime adaptation performance relative to Q-Learning, with 
comparable resource requirements.

\section{Introduction} 
\label{sec:introduction}
This paper presents a novel algorithm and a detailed application of techniques
that enable the fast and efficient use of Markov Decision Processes (MDPs) in
real-time on resource constrained Cyber Physical Systems (CPSs).  By a
resource-constrained CPS, we mean distributed systems containing components
that have embedded computers whose physical resources are constrained to be
significantly below what is available in a typical, consumer-grade desktop or
laptop computing system. These limits are typically imposed in order to reduce
the size, weight and power, as well as cost (SWAP-C) of each unit.  While this
is a relative definition of a resource-constrained CPS, what we mean in terms
of present technology are embedded systems that would typically have less than
1MB of RAM, less than 10MB of non-volatile storage, and a single-core
microcontroller with a clock speed under 100MHz.

MDPs can be used to control computing systems 
at runtime in ways that are more dynamic, robust and adaptable than alternatives.
Using MDPs, engineers can create systems 
that effectively learn and reason using models of their own system dynamics, 
observations of their own inherent limitations and effectiveness of their actions 
towards reaching application-level goals.
In this context, these systems exhibit a level of \emph{self-awareness} in their
behavior, with the ultimate design goals being continual autonomous optimization 
that leads to higher levels of runtime resiliency, robustness and efficiency.

When seeking to develop self-aware systems, researchers have recently turned to
Reinforcement Learning (RL)~\cite{sutt1998x1}, a sub-field of Machine Learning
that uses MDPs.  However, we believe that there is an important class of CPSs
that are not well served by current RL techniques.  Specifically, this class of
systems is one where some components of the system's dynamics are known at
design time, and the rest are unknown at design time or expected to be
time-varying at runtime. 

In general, RL frameworks do not try to learn the effect that control
outputs have on the system's state. Instead, they seek to use runtime
observations to find a relationship between control outputs and rewards.
In general, however, a critical part of this relationship 
is how the control output affects
the state of the system being modeled. In RL frameworks, this causality is
implicit in the modeling abstraction and not defined nor learned explicitly.
This method works well in many cases, for example in large systems where the
state dynamics are too large and complex to be considered or modeled
explicitly. However, in this work, we show that departing from this conventional
method can be useful for resource-constrained CPSs that have more
manageable state spaces. In particular, if engineers possess a priori
knowledge about how some of the control outputs might affect the system state,
it can be advantageous to codify that knowledge into the learning algorithms at
design time. The approach proposed in this work provides models and algorithms
that enable designers to exploit a priori knowledge in this way.

Motivated by this deficiency in RL, we define an alternative class of MDP-based
system modeling techniques, which we refer to as {\em Compact MDP Models} ({\em
CMMs}), and we develop CMM-based approaches as an alternative to RL for design
and implementation of adaptive CPSs. In general, we envision that certain CPSs
are better suited for RL, while others are better suited for CMM. Through this
work, we seek to explore the causes of why one of these two approaches might be
better over another for a given application, and to improve understanding to
allow designers to pick the better option.

\begin {comment}
%

\end{comment}

This paper builds upon a preliminary version in~\cite{sapi2018x2}, and 
contributes the following additional content: 

\begin{itemize}
\item A detailed survey of techniques from the literature that enable the 
compact use of MDPs on resource constrained systems. This survey leads us 
to define the class of CMM methods, which encapsulates several different 
approaches that are useful in streamlining the application of MDPs to CPS 
systems.

\item The introduction of the Sparse Value Iteration (SVI) algorithm --- 
a variation of the SPVI algorithm presented in ~\cite{sapi2018x2}. While SPVI
is a parallel processing algorithm developed for GPUs, in this work we present 
a scaled down variation that runs effectively on small, single-threaded 
Microcontrollers (MCUs).

\item A detailed example of how to apply MDP-based techniques to design a wireless sensor CPS.

\item The results of a performance simulation, illustrating the differences between
a CMM-based design approach compared to Q-Learning, a popular alternative technique from 
the Reinforcement Learning (RL) literature.

\item An empirical study of an MDP solver running on a resource-constrained MCU, including
measurements of data storage requirements, execution time and power consumption.

\item A power consumption model for an LTE-M wireless modem, derived from 
experimental lab measurements taken on a live wireless Internet Protocol (IP) 
data network. 

\end{itemize}

The remainder of the paper is organized as follows. 
We provide a cursory review of the history of techniques for controlling CPSs in 
Section~\ref{sec:background}. In Section~\ref{sec:survey}, we provide a survey of recent
advancements in CMMs. In Section~\ref{sec:method}, we present 
Structured Learning with Sparse Value Iteration, our novel method for CMM-based design.
In Section~\ref{sec:application}, 
we detail a case study of a wireless sensor CPS, and explore the 
challenges and trade-offs inherent in creating an efficient control policy. In 
Section~\ref{sec:mdp}, we illustrate how CMMs can be used to solve the design 
problem introduced in Section~\ref{sec:application} 
and compare that approach to a competing RL-based approach. Finally, in Section
~\ref{sec:simulation} we simulate the runtime performance and in 
Section~\ref{sec:implementation} we present the results of an embedded 
system implementation of the competing techniques. We conclude in 
Section~\ref{sec:conclusion} with a discussion of the results and directions for 
future work.

\section{Background and Related Work}
\label{sec:background}
The design of algorithms to control resource-constrained 
computing systems effectively at runtime has been a topic of active 
research for at least 20 years. A good survey for early work in this 
area can be found in~\cite{beni2000x2}. This survey reviews a wide 
range of techniques, including fixed threshold-based approaches, dynamic 
approaches including stochastic controllers using dynamic programming, 
as well as some guidance on how MDPs can be used in this context. 

Since the time period when that survey was written, a variety of 
approaches to this research problem has flourished over the years. 
Some researchers have sought to formulate the design challenge as a 
constrained optimization problem~\cite{kans2007x1}. Other researchers 
focused on modeling the dynamics of the system's energy consumption, and 
simplifying the control decisions to be simple threshold-based 
comparisons with respect to the energy budget (e.g., see ~\cite{liu2015x2, hsie2014x1}). 
Another popular approach has been to model the 
system as linear in the context of feedback control systems and then 
use Model Predictive Control (MPC) theory to modulate a processing duty 
cycle (e.g., ~\cite{mose2007x1, mose2010x1}).

All of these approaches were shown to be successful in their respective 
case studies, but share some common limitations when considered for use 
in other cases. For example, several of these approaches assume 
deterministic behavior from the system under test. These approaches 
model the behavior of the system in response to some actuation, and 
assume the system will always behave the same way. However, many CPSs
have some stochastic behavior, either due to complex unmodeled 
dynamics or due to being affected by an external factors that are 
difficult to predict. A second common limitation is the assumption that the 
system being controlled can be modeled as a linear system. Computing 
systems often do not behave in linear ways, and attempts at formulating 
linear approximations to non-linear behavior is limited to only the 
simplest non-linearities, which significantly constrains the overall applicability and 
generality of this approach. A third common limitation is that the 
dynamics of the system being controlled often need to be well understood 
at design time. For many computing systems whose behaviors depend 
heavily on external factors, this can be an unrealistic 
assumption. 

As efforts in this area progressed, the paradigms shifted from classical
control systems theory to various forms of adaptive algorithms, and then to
more generalized approaches that researchers have termed as self-configuration,
self-optimization and most recently, self-awareness~\cite{este2018x1}.  A wide
ranging survey of these works and organization of them into these various
self-X categories can be found in~\cite{dutt2016x1}. In that work, researchers
define self-awareness as ``attributes in a system that enable it to monitor its
own state and behavior, as well as the external environment, so as to adapt
intelligently''. Another definition can be found in~\cite{lewi2016x1}$\colon$
``self-aware computing describes a novel paradigm for systems and applications
that proactively maintain knowledge about their internal state and its
environment and then use this knowledge to reason about behaviours''.


Among the most promising directions for creating these self-awareness 
attributes in CPSs is through the use of MDPs. MDPs have shown 
success in this area because they are inherently capable of modeling 
stochastic behavior and non-linear responses, and they are also well 
equipped to deal with incomplete models and uncertainty. 

\subsection{Markov Decision Processes} 
\label{subsec:MDPs} 

MDPs provide a generic decision making framework that 
uses abstract concepts including \emph{states}, \emph{actions}, 
\emph{transition probabilities} and \emph{rewards}. Once these concepts 
are defined they are then passed to an MDP solver, which is an algorithm 
that produces an optimal policy with respect to those definitions. The 
policy is a mapping from states to actions, such that an agent using the 
policy looks up what action to take for any given state.


However, there is no consensus in the literature regarding exactly how 
to map elements of computing systems to components in the MDP framework. 
This mapping is in general left to the designer who is applying the MDP 
to solve a specific computing problem. For example, a processing system 
can be commanded to run a particular algorithm (and this can be modeled 
as a state), or that same command can be modeled as an action instead, 
or it could be modeled as both (an action that leads to a state). Also, 
the choice of granularity for these definitions is important --- e.g., 
are two invocations of the same algorithm with a slightly different 
parameter value considered two different actions, or the same action? 

There are several approaches in the literature to map elements of 
computing systems to MDP states and actions, and these different 
approaches lead to different results, with implications in both the 
final policy performance as well as how hard it is to model and solve 
the MDP. One of the earliest known applications of using an MDP to control 
resources in computing systems at runtime is~\cite{beni1999x1}. Other 
notable examples of differing approaches in the literature include a 
reconfigurable router~\cite{wei2015x2}, a reconfigurable digital filter 
bank~\cite{sapi2016x1, sapi2018x1}, a power management module for 
a microprocessor~\cite{debi2018x1}, and a smartphone scheduling program 
that synchronizes email efficiently~\cite{jung2010x1}. 


\subsection{MDP Solvers} 
\label{subsec:MDP Solvers} 

One of the first challenges associated with using MDPs is choosing what 
constitutes a state, an action and a reward. After that is decided, the 
associated MDP data structures must be stored on a computer and used as 
the inputs to the MDP solver to produce a policy. With this policy, the 
runtime decision framework consists of observing what state the system 
is in, and using that as input to the policy to determine what action to 
take. 

The classical methods to solve MDPs are algorithms known as Value 
Iteration, Policy Iteration and Modified Policy 
Iteration~\cite{russ2009x1}. All of these algorithms produce an optimal 
solution to the MDP problem, with different approaches leading to 
different implications in the execution time, power requirements and 
memory use of the solver routines. 

These classical MDP solver algorithms suffer from the same issue as most 
systems that try to reason using computations of probability 
distributions: the framework's data structures grow exponentially with 
the size of the state space. A large state space is desirable in order 
to have sufficient model expressiveness to tackle difficult decision 
problems, but this desire is at odds with the resource requirements 
needed to solve an MDP that has a large state space. The upper limits on 
memory consumption that are available on typical embedded computing 
systems can often easily be reached, before many important system 
details have been modeled. 


More specifically, the total number of elements in an MDP's 
State Transition Matrices (STMs) is ${{N_S^2}}N_A$, where $N_S$ 
and $N_A$ are the number of elements in the state space and action 
space, respectively. The STMs are the largest data structures in 
the MDP, and usually the most difficult structures to store and process, 
due to their large size. The STMs are large matrices even for 
modest choices of $N_S$ and $N_A$, and if one were to add a state 
variable with $L$ states to the state space, this addition would 
increase the size of the STMs by a factor of $L^2$. 
Besides the storage space 
and memory requirements to store large data structures, increasing the 
state space also causes the solver's execution time and power 
consumption to grow exponentially as well. 


Thus, for CPSs that operate under strict resource constraints, it is not 
enough to frame an MDP in a way that produces a well performing 
solution. There is also the practical issue of whether the solver can be 
successfully implemented on the targeted platform, and whether it can 
complete in an amount of time reasonable for the application. 

This so-called \emph{curse of dimensionality}~\cite{siga2010x1} usually 
results in limiting the use of MDPs to a mode of deployment that greatly 
hampers their usefulness: the solver is invoked only once offline, and 
then the generated MDP policy only (not the entire framework required to 
solve the MDP) is used on the target system. This scenario is suboptimal 
and limiting if the problem inputs are unpredictable, constantly 
changing, or dependent on the environment. We are interested in the more 
challenging problem of solving the MDP on demand at runtime, which 
results in a more intelligent and adaptive class of embedded systems, 
which can learn, adapt and autonomously re-optimize themselves for 
changing conditions and use cases. 

To overcome the limitations of MDPs with this goal in mind, two main 
approaches have been pursued: RL and CMMs. 
RL essentially tries to arrive at the policy 
without explicitly modeling all of the MDP components or invoking 
a solver. On the other hand, CMMs are approaches that do define all of the MDP components
and invoke a solver, but do so via algorithmic optimizations
that significantly reduce computational requirements. These two
alternatives are sometimes referred to as model-free and model-based 
RL, respectively. Henceforth in this paper, when we write ``RL'', we refer to 
model-free RL unless otherwise stated. 

We discuss these two categories --- RL and CMM --- of techniques in 
Section~\ref{subsec:RL} and Section~\ref{sec:survey}, respectively. 

\subsection{Reinforcement Learning} 
\label{subsec:RL} 

\begin{figure} 
\centering 
\includegraphics[width=3in, angle=0]{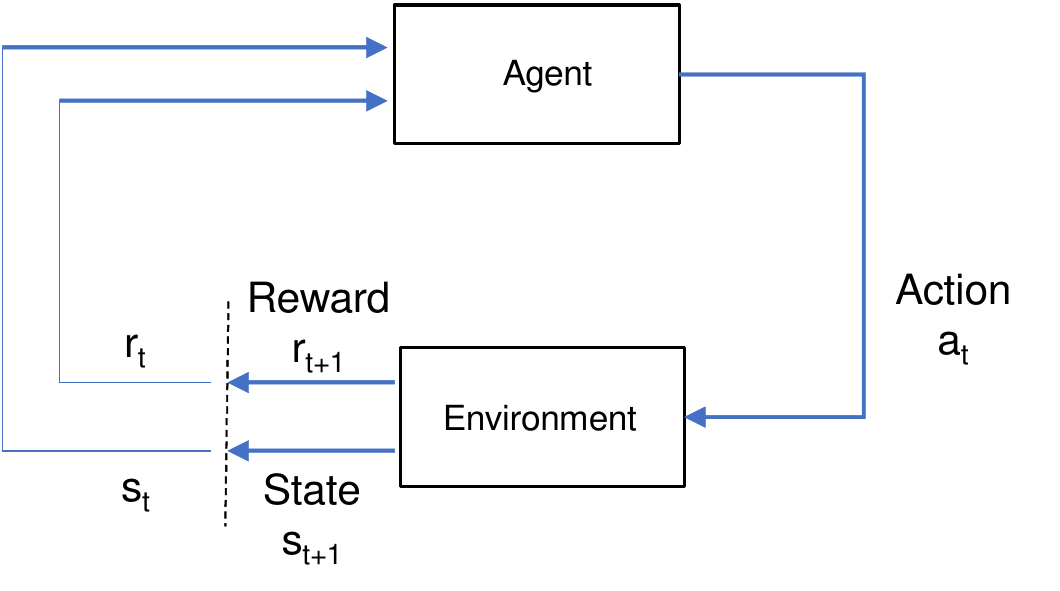} 
\caption{Block diagram of reinforcement learning paradigm.} 
\label{fig:rl} 
\end{figure} 

Reinforcement Learning (RL) ~\cite{sutt1998x1} is an area of machine learning 
that enables systems to formulate optimal decision policies using 
observations of rewards that are received for previous decisions at 
runtime. These techniques use MDPs as the framework for formulating the 
decision problem, but seek to learn the optimal policy directly using 
observations of rewards in response to decisions, rather than 
through the explicit definition of all of 
the MDP components followed by invocation of an MDP solver. 

More specifically, an RL framework typically contains the top level 
block diagram shown in  
Figure~\ref{fig:rl}. The learning takes place by some agent, which is 
responsible for selecting an action out of a set of actions, given a 
system state. This selection is usually done in a discrete-time setting 
and iterated at a fixed rate. Each selected action (in a given state) 
leads to some consequence in the environment, and 
that causes it to transition into a 
new state in the state space. The selected action and transition to that 
new state are associated with a scalar reward, which is fed back to the agent 
(positively or negatively). The agent in turn considers the reward it 
has been given along with the new state that resulted, and again selects 
the next action, repeating indefinitely. 

As mentioned in the previous Section, these techniques are 
sometimes referred to as model-free learning. Model-free learning techniques 
possess the advantage of completely bypassing the need to maintain large STMs, 
and run computationally intensive MDP solvers.

However,
this advantage comes at a cost, as the consequences of all actions taken in
all states have to be learned and constantly validated, 
even those that are constant and known a priori at design time. 
This learning comes at the cost of occasionally having to make random decisions at 
runtime to explore the effect of alternative decisions\cite{tijs2016x1}.
This cost is a central drawback of model-free techniques, and is associated 
with a complex trade-off called the {\em exploration versus exploitation 
trade-off}~\cite{sutt1998x1}.

One popular RL technique that has shown promising results is 
Q-Learning~\cite{debi2018x1, hsu2014x1, liu2010x1, yue2012x1}. 
In Q-Learning, a scalar value ``Q'' is assigned to 
each action in each state, and referred to as the Q-Function. This 
function represents the average future rewards
that can be expected by taking a given action in a given state. The 
Q-Learning method continually learns and updates this Q-Function 
using a simple technique called the method of Temporal Differences (TD)
~\cite{moda2012x1}, 
and uses it to formulate an optimal control policy that is based entirely 
on the system and environment involved. As the environment and the system's
dynamics change at runtime, the policy changes with it. 

In one example~\cite{debi2018x1}, an Adaptive Power Management (APM) 
hardware module using Q-Learning was used to put a microcontroller in 
and out of low power states, and resulted in a learning controller that 
managed power transitions better than an expert user. In another 
example~\cite{hsu2014x1}, Q-Learning was used to optimize the 
throughput of an energy harvesting wireless sensor node while meeting 
challenging constraints. 

In the next section, we give an overview of CMM methods, which can be viewed as 
model-based RL methods that are designed with an emphasis on streamlining 
computational efficiency. We survey recent advancements in this area that 
result in performance on par with Q-Learning.


\section{Survey of Compact MDP Models}
\label{sec:survey}
As mentioned in the previous section, the deployment of MDPs in 
resource-constrained systems has typically been limited to usage modes 
where the MDP modeling and solving are done offline, and only the 
resulting policy is stored in the runtime system. 
The goal of moving the MDP model and solver into 
the runtime system by mitigating the longstanding barriers has been a 
common goal among many researchers over the years, resulting recently in 
very creative and effective techniques. We refer to this 
category of approaches as compact MDP models (CMMs), given 
that they provide a smaller or computationally optimized representation of the 
system in question than compared to a direct implementation of the MDP's 
data structures. 
The effectiveness of these recent developments, especially when applied in 
combination with one another, leads us to question the conventional approach 
of limiting consideration to model-free RL approaches in the implementation 
of resource-constrained, MDP-based systems.

Some of these CMM techniques seek to reduce the storage size of the 
MDP's data structures by exploiting some structural component embedded 
within the MDP (e.g., see~\cite{sapi2018x2, bout1995x1, hoey1999x1, jons2006x1, lin1995x1}). 
Other 
techniques involve modeling approaches that reduce the MDP state space 
via generalization and abstraction of system dynamics (e.g., see~\cite{sapi2016x1, sapi2018x1}). 
Another approach has 
been to keep algorithms and data structures as is, and take advantage of 
recent advancements in parallel processing using embedded GPUs, for 
example ~\cite{ruiz2015x1, sapi2018x2}. 

In our case study, we utilize three CMM techniques: factorization, exploitation
of sparsity, and transition states, and we show the significant advantages that
they provide when compared to both a direct classical MDP implementation and a
competing model-free method. In the remainder of this section, we present an
overview of four important CMM techniques from the literature, including the
three techniques that are used in our case study. One of the techniques
surveyed here --- hierarchical decomposition --- is not used in our case study,
but still presented here due to its importance and widespread use in the field.

\subsection{Factorization}
\label{subsec:factorization}

In MDP problems, the state $s \in \stsp$ is constructed to model the problem
the MDP is being applied to. Often this results in the state being an
instantiation of a discrete multivariate random variable $\underline{Z} = (Z_1,
Z_2, \ldots, Z_{N_Z})$, with each variable $Z_i$ taking on values in
$\mathit{DOM}(Z_i)$, where $\mathit{DOM}(X)$ represents the set of admissible
values of the random variable $X$. A state is a set of instantiations of the
$N_Z$ random variables, and can be written as a vector $\underline{z} \in
\mathit{DOM}(\underline{Z})$. The size of the state space is defined by the
cardinality of this set, which we denote as
$\mxcard{\mathit{DOM}(\underline{Z})}$. As a result, each row of each
transition matrix for an MDP has width $\mxcard{\mathit{DOM}(\underline{Z})}$,
and describes the probability of reaching all possible combinations of the set
of variables $(Z_1, Z_2, \ldots, Z_{N_Z})$.

We refer to MDPs with this kind of formulation as having a multivariate state
space. When an MDP's STMs  are stored in a structured way that uses knowledge of
the causal relationships between these state variables to reduce storage size, the MDP 
is said to be \emph{factored}.
To the best of our knowledge, the earliest publication detailing the concept of 
factored MDPs is Boutilier et al.\ in~\cite{bout1995x1}. These researchers
proposed factored MDPs as a
method for compact representation of large, structured MDPs.
Through application and empirical observations, we have found that this method 
can effectively reduce STM storage size considerably. However, it requires a specific
conditional probability structure to be present in an MDP, and the data
structures must be created by hand with specific knowledge of the exact
structure. 
This can require a subject matter expert in the loop
anytime a transition probability changes, which can 
complicate runtime autonomous solving of an MDP that changes
over time in unknown ways.  In general, this requirement can be problematic if 
the underlying structure is not fully understood. We acknowledge the effectiveness 
of the technique, but also the fact that it cannot always be used.

We applied this technique in ~\cite{sapi2018x2} where we used a factored MDP to 
create a control policy for a reconfigurable digital filter bank. Factoring the 
MDP resulted in a reduction of the number of elements in the STM from 
$121.8x10^6$ to $66.3x10^3$. This is a very important reduction, because a matrix
with 66k elements can realistically be stored in a low power MCU, whereas one 
with 121M elements typically cannot. 

With the goal of creating a custom solver that works directly on a factored MDP, 
Hoey et al.\ ~\cite{hoey1999x1} detail an algorithm similar to Value Iteration that can solve 
MDPs in factored form using Algebraic
Decision Diagrams. This approach shows good results in
taming the curse of dimensionality, but imposes the same restrictions
as~\cite{bout1995x1} and thus has the same limitations. Additionally,
the state transition matrices must be manually converted into
tree-shaped conditional probability structures, which we found to be considerably 
difficult and time-consuming.

\subsection{Hierarchical Decomposition}

In ~\cite{jons2006x1}, Jonsson and Barto present an algorithm that performs 
hierarchical decomposition of factored MDPs into smaller subtasks to help 
alleviate the growth in complexity that follows from a modestly-sized state space. 
This approach can
be effective, but also requires a priori knowledge of the causal structure
within the MDP. This knowledge can be very difficult or impossible to know
for many MDPs, a shortcoming identified by the authors themselves. 
Also, this method requires that the MDP have this decomposability 
property, which is not always the case.

In a similar spirit, Lin and Dean~\cite{lin1995x1} present a method to solve a
large MDP by first decomposing the state space into regions, determining
actions to take within those regions, and then using novel approaches to
combine the resulting sub-policies into an overarching policy that solves the
original, large MDP. The authors note in this work that the decomposition must
be done a priori by a domain expert. This decomposition is not guaranteed to be 
feasible, and when it is feasible, it can be very difficult to perform.

\subsection{Sparsity}
\label{subsec:sparsity}

Another approach in reducing the computational requirements of MDPs is 
the exploitation of the sparsity typically found in the MDP data structures. 
In this context, we refer to sparsity as a high percentage of
zero-valued elements in the MDP STMs. A 
discussion on the reasons why sparsity typically results in MDPs can be found
in~\cite{sapi2018x2}. In~\cite{wijs2016x1}, Wijs et al. present a promising method to decompose MDPs into
subgraphs, exploiting sparsity on GPUs. However, the method is presented in the
context of model checking for formal methods in software engineering. More
research is required to incorporate this method into an MDP solver. 

In ~\cite{sapi2018x2}, we introduced Sparse Parallel Value Iteration (SPVI), an
MDP solver algorithm that exploits sparsity through a three-step process: 
1) an algebraic manipulation of the
MDP components is performed that combines the multiple STMs into a single large
matrix, 2) the large matrix is stored in a sparse matrix format, and 3) a 
specialized form of the Value Iteration algorithm is used to operate on the MDP
directly in this transformed state. 
The examples in~\cite{sapi2018x2} detailed significant
improvements in both storage requirements and solver computation time, however that
work also exploited parallelism in GPUs to achieve this result. In Section~\ref{sec:implementation},
we show that the sparse linear algebra techniques are also quite effective when running
on a resource constrained single threaded MCU, not just on the much more powerful processing
environment (including hundreds of parallel threads) provided by GPUs. When running in a
single threaded CPU, we no longer have any parallel processing and thus refer to the 
algorithm more simply as Sparse Value Iteration (SVI). We believe that our SPVI and SVI 
solvers are the first reported MDP solver implementation (either CPU- or GPU-based) 
that exploit sparsity to  make significant performance gains in both storage 
requirements and computation time.

\subsection{Transition States}
\label{subsec:transition-states}

In~\cite{sapi2016x1}, we introduced the concept of \emph{transition states}, which is based on a
similar concept initially proposed in~\cite{beni1999x1}. We expanded and elaborated on the 
initial concept in both~\cite{sapi2016x1, sapi2018x1}.

In our design context in this paper, the system being controlled contains many discrete states.
Depending on the level of modeling and decision making that is desired, as much as tens of thousands 
of states or more could be considered relevant. 
In general, the modeling process involves making design-time
decisions as to what level of detail is modeled in the MDP, for each of the system 
characteristics and dynamics. More fine-grained detail allows for a more precise model, but 
this leads to a large state space and the computational challenges associated with that. 

The concept of transition states allows for significant reduction of the state space to 
occur when the system passes through a large set of states for a limited time,
and the only relevant detail is when the system enters and exits the set as a whole. 
In some way, these trajectories need to be modeled in the MDP.
By utilizing the concept of transition states, large groups of fine-grained states are 
abstracted away into a single coarse-grained state referred to as a transition state. 
While the system is actually in one of the many fine-grained states, the system 
is modeled as being \emph{in transition} through the single coarse-grained 
transition state. The transition state encapsulates a set of discrete states that are 
present but not relevant to the decision process being designed. The only transition 
probability needed is derived from an estimate of the expected time until the system
leaves the set of states. A discussion on how to apply this to stochastic dynamics, 
and some of the trade-offs associated with this concept can be found in
~\cite{sapi2016x1, sapi2018x1}.

\section{Method}
\label{sec:method}
\subsection{Structured Learning}

Creating an MDP model on a computing system consists of defining the 
states and actions, the STMs, and the reward function for the given
decision problem and its environment. 
The STMs are $N_A$ stochastic matrices, each of size $N_S$ by $N_S$ (one matrix
for each action). Each STM defines the probability of 
transitioning from the current state to any one of the possible other 
states, given an action. We generally write this as a discrete 
conditional probability distribution as in 
Equation~\ref{eq:stm_general}:

\begin{equation}
\label{eq:stm_general}
p(s^{(n+1)} | s^{(n)}, a^{(n)}), \forall s \in \stsp, a \in \acsp,
\end{equation}

\noindent which gives the probability of the system transitioning to state 
$s^{(n+1)}$ at time index $n+1$ given that it was in state $s^{(n)}$ and 
action $a^{(n)}$ was selected at time index $n$. The process of 
instantiating the STMs is the allocation of storage for 
${{N_S^2}}N_A$ numerical quantities and assigning them values from 0 to 
1. Given this viewpoint, we group the methods in the literature into two 
categories: the model-based approach where all $N_A{N^2_S}$ terms are 
defined a priori and treated as constants, and the model-free 
approach, where none of the ${{N_S^2}}N_A$ are defined and in fact 
storage for them is never even allocated. 

In this paper, we view these as two extremes of a continuum that has many 
other options. We propose a blend between the two, where some of the 
STM terms are assumed to be known a priori, and others are not. 
More specifically, we define: 

\begin{equation}
\label{eq:structured_learning_stm}
\begin{split}
p(s^{(n+1)} | s^{(n)}, a^{(n)}) \in \{\Gamma \cup \hat{\Theta}\},\\ \forall s \in \stsp, a \in \acsp,
\end{split}
\end{equation}

\noindent where we imply that all of the probability values in the STMs
come from one of two parameter subsets $\Gamma$ and $\Theta$. The set 
$\Gamma$ is the set of STM entries that are fully known a priori 
and can be set to a fixed value at design time. The set $\hat{\Theta}$ 
contains the remaining matrix elements; they are either not known a priori or are 
expected to be time-varying at runtime. We use $\hat{\theta}$ to 
denote the latest value of a running of estimate of the true value, for each
parameter $\theta \in \Theta$.

Rather than taking the entire STM as either constant, or completely 
unknown, we adopt a flexible middle ground and assume it to be partially known 
and partially unknown. In this way, the system model has some parts of 
it that are fixed, and other parts that are assumed to 
change over time. Then, the runtime adaptation process consists of 
learning only the set of parameters $\hat{\Theta}$, rather than the 
entire STM. In this way, the model contains a mix of some 
predetermined structure from $\Gamma$) and some online learning (from 
$\hat{\Theta}$). We refer to this approach as Structured Learning. 

Structured Learning allows us to restrict how much effort is spent trying to 
learn unknown parameters, and results in higher overall awareness and 
adaptation performance for a certain class of CPS devices, as will be 
demonstrated in our case study. The advantage comes from being able to direct 
the system's learning efforts to be focused on the relevant parts of the 
problem, and prevent redundant attempts to constantly question and re-visit 
assumptions about the system that a system designer knows will never change.

\subsection{Temporal Difference Equations} 

The Structured Learning method defined above relies on a continual 
online learning of parameters. For this, we use a very simple technique 
prevalent in RL: the use of weighted averaging through Temporal Difference (TD) 
equations, with the central concept shown in Equation~\ref{eq:td}:

\begin{equation}
\label{eq:td}
\hat{\theta}^{(n+1)} = \hat{\theta}^{(n)}\cdot(1-\alpha) + \theta^{(n)}\cdot \alpha,
\end{equation}

\noindent where $\theta^{(n)}$ is an observed value of one of the parameters 
$\theta \in \Theta$ at timestep $n$, $\hat{\theta}^{(n)}$ is the value 
of the running estimate of $\theta$ at timestep $n$, and $\alpha$ is a 
learning rate parameter which controls how sensitive the running 
estimates are to individual observations. The method essentially consists of 
performing a low-pass filtering or smoothing operation on observed 
values, and thus maintaining a running estimate that tracks the latest 
observed values for a given parameter. 

As the observations change over time, the running estimates track the 
changes while also reducing the effect of statistical outliers. Using 
TD, the Structured Learning method can ingest the latest observations of 
each of the parameters $\theta \in \Theta$, compute the running 
set of estimates for each $\hat{\Theta}$, and combine them with the constant set 
$\Gamma$ to assemble the fully populated, partially time-varying STMs 
at any timestep. 

\subsection{Sparse Value Iteration (SVI)}
\label{subsec:svi} 

In Structured Learning we instantiate the full set of STMs. In order to 
obtain a control policy from this, we must invoke an MDP solver, and for this 
we use a modified version of the Value Iteration algorithm. 
In Value Iteration, a real number (or value) $V(s)$ is associated with each state $s$.
This mapping is known as the Value Function. The value $V(s)$ represents the
expected reward that can be obtained from state $s$.  The Value Function $V$ is
derived by using the iterative procedure shown in Equation~\ref{eq:vi_backup},
which starts out assigning a value of zero for each state and then
incrementally converges from that to an optimal Value Function.  Once
sufficient iterations are performed, the optimal Value Function is known and
the optimal MDP policy can be obtained trivially from it. This process of
deriving the Value Function is a form of dynamic programming~\cite{bell1984x1}:

\begin{equation}
\label{eq:vi_backup} 
\begin{split}
V^{0}(s_i)&=0 \\
V^{n}(s_i)&=\max_{a \in \acsp}\{R(s_i,a) + \beta\sum_{s_j \in \stsp}[P(s_j|s_i,a)V^{n-1}(s_j)] \}.
\end{split}
\end{equation}

In Equation~\ref{eq:vi_backup}, ${V}^{n}(s_i)$ is the approximation to the
Value Function in state $s_i$ at loop iteration $n$, $\stsp$ is the discrete
state space, $\acsp$ is the discrete action space, $R(s_i,a)$ is the reward
for the state-action pair $(s_i,a)$, $\beta$ is a scalar discount
factor, and $P(s_j|s_i,a)$ is the probability of transitioning from state $s_i$
to state $s_j$ after taking action $a$. Arranging the conditional probabilities
in a matrix with $s_i$ as rows and $s_j$ as columns gives the State Transition
Matrix (STM) for action $a$.

Next, we rewrite Equation~\ref{eq:vi_backup}
into in Equation~\ref{eq:vi_backup_vectors}, using matrix and vector
representations. 

\begin{equation}
\label{eq:vi_backup_vectors}
\underline{V}^{n}=\max_{a \in \acsp}\{\underline{R}+\beta\cdot \boldsymbol{M} 
\cdot\underline{V}^{n-1}\}.
\end{equation}

\noindent Here, $\underline{R}$ represents the reward function for each state
and action flattened into a length $N_S N_A$ column vector, where $N_S$ and 
$N_A$ are the number of elements in the state space and action space, 
respectively; $\beta$ is (as defined previously) the discount factor, which is a scalar;
and $\boldsymbol{M}$ represents the 
vertical concatenation of all $N_A$ of the $N_S \times N_S$ transition matrices 
into a single $(N_S N_A) \times N_S$ matrix.

Through execution profiling, we observed consistently that a large portion of
the total computation time in Value Iteration is spent multiplying the
$\underline{V}^{n-1}$ values by the transition probabilities. In other words, a
large portion of the computation time in Equation~\ref{eq:vi_backup} 
is spent performing the summation loop over $s_j \in
\stsp$, which needs to be repeated (${N_S}{N_A}$) times for each iteration.
Equivalently, in
Equation~\ref{eq:vi_backup_vectors}, the majority of the time is spent
performing the large matrix-vector multiplication $\boldsymbol{M}
\cdot\underline{V}^{n-1}$.  


Under the assumption that the matrix $\boldsymbol{M}$ is sparse, we conclude
that the majority of the computation time in Value Iteration solvers is spent
multiplying a large sparse matrix by a vector. In other words, much of the time is
spent multiplying elements by zero and then summing those zeros to other zeros.
SVI exploits the same principle as all sparse linear algebra software
libraries --- that an operation that is guaranteed to produce a known result
(zero), can be skipped altogether resulting in a performance improvement in
time, memory use, and power consumption. By replacing linear algebra operations 
with operations that are specifically optimized for sparse matrix-vector algebra, 
we can achieve a significant improvement in performance gain beyond the current 
state of the art in MDP solvers.

A sparse matrix format is a data structure that represents a matrix. However,
instead of the standard approach for matrix storage (a serialization of each
element in the matrix regardless of its value), a sparse matrix structure
contains an array of just the non-zero elements, along with two other arrays
that indicate where those elements are located in the matrix. The format
implicitly assumes that elements not specified are zero by default. In this
form, a matrix can be represented with no loss of information, and if the
sparsity of a matrix is high, then the sparse representation can be much
smaller than the matrix stored in a standard (fully serialized) format.
Correspondingly, multiplying a sparse matrix by a vector can be much faster
and memory-efficient if the sparsity is high. 

A pseudocode description of the SVI algorithm is shown in
Algorithm~\ref{alg:svi}.

\begin{algorithm}
\caption{Sparse Value Iteration (SVI).}
\label{alg:svi}
\SetAlgoNoLine
\LinesNumbered
\DontPrintSemicolon
\KwIn{$\stsp$, $\acsp$, $R(s_i,a)$, $P(s_j|s_i,a)$, $\beta$, $\tau$}
\KwOut{$\underline{\pi}$}
Compute $K_{\mathit{NZ}}$, the number of non-zero elements in $\boldsymbol{M}$\;
Allocate memory for $\boldsymbol{M}_s$, a sparse matrix sized for $K_{\mathit{NZ}}$\;
\For{each element $z$ in $\boldsymbol{M}$}{
if $z \neq 0$ add it to $\boldsymbol{M}_s$\;
}
$\underline{V}^{0} \leftarrow \underline{0}$\;
$n \leftarrow 0$\;
\Repeat{$\Delta < \tau$}{
$n \leftarrow n+1$\;
$\underline{T} \leftarrow \mathit{SPARSE\_MULT}(\boldsymbol{M}_s, \underline{V}^{n-1})$\;
$\underline{Q} \leftarrow \mathit{SAXPY}(\beta, \underline{T}, \underline{R})$\;
$\underline{V}^{n},\underline{\pi}^{n} \leftarrow \mathit{MAX\_REDUCE}(\underline{Q})$\;
$\Delta \leftarrow \mathit{INF\_NORM}(\underline{V}^{n}, \underline{V}^{n-1})$\;
}
$\underline{\pi} \leftarrow \underline{\pi}^{n}$\;
\end{algorithm}

\subsubsection{State Transition}

The computation of the product $\boldsymbol{M} \cdot \underline{V}^{n-1}$ 
in Equation~\ref{eq:vi_backup_vectors} is
efficiently implemented in SVI using a sparse Matrix-Vector multiplication. The
sparsity in the transition matrices is exploited by the conversion of the large
and sparse $\boldsymbol{M}$ to a much smaller, densely packed
$\boldsymbol{M}_s$ on lines 1 through 5. Then, a multiplication of the
sparse matrix by a vector is performed using a subroutine denoted by
$\mathit{SPARSE\_MULT}$ in Line 10. This subroutine is a
standard sparse matrix-vector multiplication.

The sparse matrix $\boldsymbol{M}_s$ is created in the Compressed Sparse Row Matrix
format. This conversion only needs to be performed once at initialization.
Details on this sparse matrix format, as well as a thorough analysis of the
history and performance advantages of performing sparse Matrix-Vector
multiplications can be found in~\cite{fili2017x1}. 

Next, the discount factor and rewards need to be applied. After computing the
product $\boldsymbol{M}_s \cdot \underline{V}^{n-1}$, the remaining steps can
be implemented by scaling the product by a scalar $\beta$ and then adding it to
the vector $\underline{R}$. This is a common operation referred to as a
Single-Precision A $\underline{X}$ plus $\underline{Y}$ (SAXPY), and is very
efficiently implemented in most linear algebra packages. We denote this 
subroutine here as $\mathit{SAXPY}$ in Line 11 of Algorithm~\ref{alg:svi}.

\subsubsection{Action Selection}

The $N_S$ elements of the Value Function $\underline{V}^{n}$ and policy
$\underline{\pi}^{n}$ are computed from $(N_S N_A)$ elements of the
$\underline{Q}$ vector, which constitutes the selection of an optimal action
for a given state. This computation is invoked in Line 12
of Algorithm~\ref{alg:svi}. The subroutine computes the maximum
value (and the action associated with it) from a subset of $\underline{Q}$, 
striding across the vector only on the elements associated with each state.  

\subsubsection{Stopping Criteria}

In order to evaluate the stopping criteria for SVI, the infinity norm of the
incremental approximations to the Value Function must be computed. This
operation is represented by Lines 13 and 14 of Algorithm~\ref{alg:svi}.

\section{Application}
\label{sec:application}
In this section we detail a specific type of CPS, which we use as our 
case study for the remainder of the paper. The CPS is an embedded system 
with constraints on its size, weight and power (SWaP), containing the 
physical components shown in the following list. 

\begin{itemize} 
\item A sensor and/or actuator to interact with the physical environment. 
\item A wireless modem used to provide Internet access to the system. 
\item A low-power Microcontroller Unit (MCU) 
executing a program that controls the sensor and/or actuator as well as 
the wireless modem. 
\item An energy source that is used to power the 
system. The source can be a battery that needs to be replaced 
periodically, or an energy harvesting source (such as a solar panel 
paired with a rechargeable battery). 
\end{itemize} 

This type of CPS is expected to exist as one instance of a plurality of 
identical nodes in an installed base, and the nodes are 
connected to an application server via an Internet connection. We seek 
to empower each node with the ability to optimize its performance for 
its own specific environmental conditions, rather than centralizing all 
of this optimization responsibility in the cloud. This approach is 
advantageous in terms of scalability and also reducing communications 
overhead. 

The MCU runs an application-specific program that utilizes the sensor 
and/or actuator to interact with its physical environment. The 
application is typically multi-modal, in the sense that it does not 
always do exactly the same application-level operations at all times. The 
application may, for example, have two modes such as 1) the normal 
sensor and actuator operating mode, and 2) a firmware update mode where 
security patches and product updates are routinely downloaded to the CPS node. It 
is expected that the application could have more than two modes. 

The CPS node's power source is typically very limited in its capacity. 
In the case of a solar panel, this may be due to a desire to keep the 
solar panel cost low and the size small. In the case of a battery, this 
may be due to a desire to maximize battery life in order to reduce how 
often the battery needs to be recharged or replaced. Regardless of the 
power source, the CPS node's application is generally tasked with 
carrying out its functions in the most energy efficient manner possible 
due to limitations in its energy source. 

In order to maximize battery life, a low-power MCU is used, rather 
than a general purpose computing system. As a result of this, the CPS 
node will be limited in its CPU frequency, RAM size, and non-volatile 
storage capacity. This fact becomes important when considering the use 
of computationally intensive control algorithms. A more involved program 
requires more computing resources, which in turn requires the use of a 
more capable computing system that consumes more power (even while 
idle). Thus, a thorough consideration of control algorithms must analyze 
not just the control performance, but also the computational 
requirements that are needed to deploy it on a self-contained and 
resource-limited MCU. 

The MCU is also tasked with controlling the wireless modem to enable 
communication typically with another Internet-connected device such as 
another CPS node or a cloud-based application server. In this case study, we 
focus on one specific form of wireless network that is very common at 
the time of this writing: LTE-M (LTE for Machines), also known as LTE 
Cat-M1~\cite{dawa2016x1}. This wireless protocol is a subset of the full LTE 
protocol, the wireless network that makes up the majority of cellular 
Internet connections at the time of this writing. LTE-M is a reduced 
form of the full protocol, and is designed specifically for resource 
constrained devices. In the following section, we describe the power 
profile of an LTE-M modem in detail in order to illustrate the power 
management considerations an MCU controller must balance. 

\subsection{LTE-M} 
\label{subsec:ltem}

The LTE-M modem is usually the largest consumer of power in the 
processing and communication subsystems. When actively 
communicating with a cell tower, an LTE-M modem's average 
power consumption can be as much as 650 milli-Watts. This quantity is 
very high relative to the consumption of other components in the CPS 
node. As a result, leaving the LTE-M modem powered on and connected to a 
cell tower continuously is usually not a feasible option for CPS nodes, 
from the point of view of maximizing the lifetime of a limited energy 
supply.  As a result, the MCU must turn 
the LTE-M modem on and off strategically in order to stay within a 
limited energy budget. 

\begin{figure} 
\centering 
\includegraphics[width=4in, angle=0]{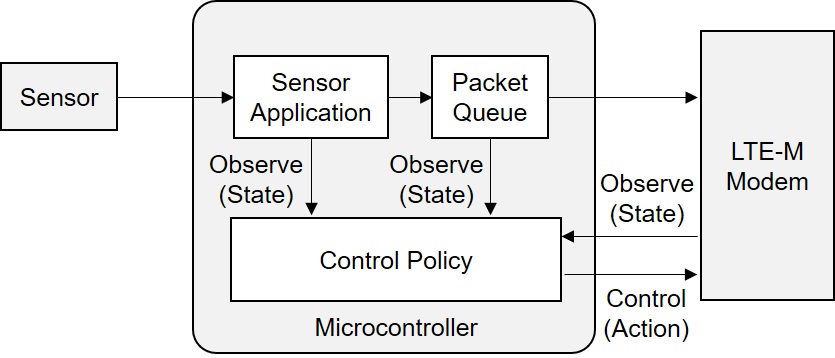} 
\caption{Block diagram of wireless sensor CPS.}
\label{fig:block_diag} 
\end{figure}

In order to fully understand this power management challenge, we 
obtained a Sequans Monarch VZM20Q LTE-M modem and a cellular 
data plan for the Verizon Wireless LTE-M network in the United States. 
This is a live Internet Protocol (IP) network with nearly nationwide 
coverage, and effectively provides a mobile Internet connection that can 
be accessed from almost any location by an MCU. 
In our work, we focus on the upstream flow of 
information. That is, the collection of data through a sensor and the 
transmission of that sensor data up to a cloud-based server. The resulting 
information flow within the CPS node is outlined in 
Figure~\ref{fig:block_diag}, where the arrows interconnecting the 
blocks represent the flow of information from one component to another.

\begin{figure} 
\centering 
\includegraphics[width=4in, angle=0]{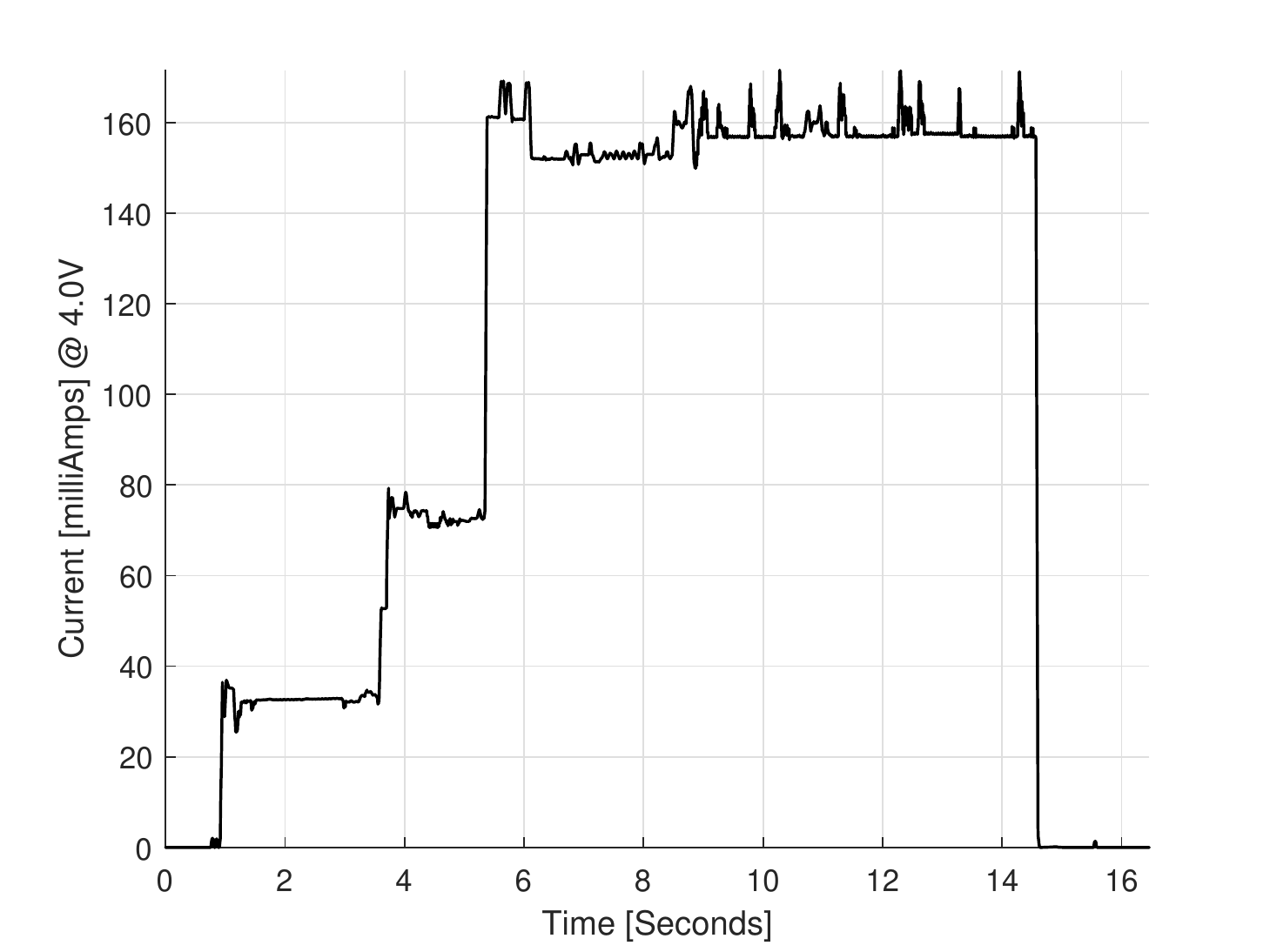} 
\caption{Current consumption of LTE-M modem during data transfer.}
\label{fig:ltem-power} 
\end{figure}

Due to the LTE-M power profile, a feasible use case for a system like 
this would be to keep the modem powered off by default, 
periodically turn it on to exchange data with the Internet and then 
power it back down to conserve energy. To model this use case, we 
controlled our Sequans Monarch LTE-M modem from a test application on a 
laptop computer that powered the modem on, connected to the nearest 
Verizon Wireless cell tower, transmitted some information (representing 
a sensor reading) from the test application to a cloud-based server, 
then disconnected the cell tower connection, and finally powered down 
the modem. During multiple runs of this test, we measured the power 
consumption of the LTE-M modem. A typical current versus time trace of 
the energy required by the modem to perform this procedure is shown in 
Figure~\ref{fig:ltem-power}. The trace is a time-series of current draw 
in milli-Amperes at a fixed voltage of 4.0 Volts, and thus instantaneous 
power and total energy for the transaction can both be computed from the 
data.

Additionally, we varied the quantity of data that we transmitted to the 
server on each test run. Specifically, we used a packet size of 256 
Bytes per packet and varied how many of these fixed-sized packets were 
transmitted. These measurements helped us quantify the approximate 
energy consumption of the LTE-M modem as a function of the amount of 
data transmitted. This allowed us to create an approximate model of the 
modem power consumption, as a function of the quantity of packets 
transmitted. Our derived model is given by Equation~\ref{eq:energy_per_packet}:

\begin{equation} 
\label{eq:energy_per_packet} 
\begin{split}
E_{TX}(N_T) = c_1 + c_2 ({N_T}-1) \,[\mathit{Joules}]\\ {N_T}\geq1, c_1=6.62, c_2=1.55,
\end{split}
\end{equation}

\noindent where $E_{TX}$ is the energy consumption in Joules and $N_T$ is the 
number of packets. This equation illustrates a defining characteristic 
of this type of wireless connection: the energy overhead of powering the 
modem on and connecting to a cell tower can be significantly higher than 
the incremental cost of transmitting a packet once the connection is 
active. Thus, if optimizing strictly for energy efficiency it is 
advantageous to queue up multiple packets before powering on the modem, 
and then transmit the queued packets together. This approach, 
however, is the opposite of what should be done if optimizing for 
transmission latency. This is a fundamental trade-off of controlling 
the LTE-M modem in this CPS node. 

The power control challenge for a system like this consists of 
implementing an algorithm that strategically determines when to allow 
new sensor data to accumulate in the sensor, versus when to invoke a 
communication event (which would produce a power consumption 
profile similar to the one 
shown in Figure~\ref{fig:ltem-power}) that transfers all packets in the 
queue up to the cloud-based server. This decision problem involves a 
trade-off of energy efficiency versus communication latency. 

A trivially simple policy is one where any time a new packet arrives in 
the queue, it is immediately transmitted up to the cloud. 
In fact, this could be considered a default case where no control policy 
analysis had occurred, and the MCU program was designed simply to 
transmit whenever a new packet exists. However, this is still a control 
policy and we consider it as such. This policy will have a low energy 
efficiency (which is undesirable) and low communication latency (which 
is desirable). 

In another example, a controller could wait until a fixed number of 
packets accumulate in the queue before doing a batch transfer of all 
accumulated packets up to the 
cloud. This policy would have higher energy efficiency but also higher 
communications latency, as some packets might sit in the queue for some 
time before being sent to the server. Ultimately, the CPS node's 
application will dictate what type of latency is desirable
for the system, and it should be made as efficient as possible while 
meeting the specified latency goals.

The policies described above are both very simple ``fixed threshold''
policies. These are easy to implement and analyze for the dynamics 
described thus far. However, in a real-world CPS deployment, the system 
control dynamics are likely to be more complicated and time-varying than 
has been described so far. Additional complexity arises from the 
following phenomena: 

\begin{itemize} 
\item The application is likely to be multi-modal and 
thus the rate of packet generation within the CPS node is in general 
time-varying. 
\item The number of different modes that the application may contain could be 
much more than two. 
\item The modem connection time 
and power consumption are also time-varying due to the 
physical mobility of the CPS node relative to the fixed 
location of the cell tower as well as the presence of network congestion from 
other LTE-M users within the same cell during periods of heavy usage.
\end{itemize} 

When these real-world factors are modeled, it becomes advantageous to 
consider more involved control policies that can monitor and adapt to 
the time-varying conditions which may not be precisely modeled prior to 
a system deployment. For this reason, we consider two approaches that 
contain online learning capabilities specifically to self-optimize 
continuously at runtime, adapting to the time-varying nature of the CPS 
node's environment. This feature avoids having to understand and anticipate
all of the runtime conditions ahead of time, which for practical 
purposes is an infeasible requirement for real-world deployments that have
high numbers of nodes.

In Sections~\ref{sec:mdp} through~\ref{sec:implementation}, we first describe 
multiple options for controlling the
CPS node in our case study and then compare the performance of the 
policies with regard to the efficiency versus latency trade-off in 
simulation. Afterwards, we implement the competing options on a resource 
constrained MCU and detail the resulting computational requirements and 
deployment feasibility.

\section{MDP-Based Control}
\label{sec:mdp}
In this section we detail two approaches to create control policies
for the CPS node introduced in Section~\ref{sec:application}, and then 
compare their performance using simulation. Both approaches use 
discrete-time MDPs to generate a control policy, which determines when 
to power the LTE-M modem on and off. The controllers are both designed 
with the assumption that several of the system's characteristics and 
dynamics are time-varying and uncontrollable. Thus, both controllers 
employ some form of learning, in that the policy that each controller employs is 
continually updated to reflect the time-varying changes in both the 
system it is controlling and the environment that the system interacts 
with at any given time. 
The two controllers can be viewed as being two different realizations of 
the agent component in the framework of Figure~\ref{fig:rl}. We refer to 
the two methods as: 1) the Structured Learning controller and 2) the 
Q-Learning controller. The Structured Learning controller is a novel 
approach described for the first time in this paper, and the Q-Learning 
controller is a well known technique in the RL literature (e.g.~\cite{sutt1998x1}). 


\begin{table}[]
\caption{Comparison of two MDP-based controllers.}
\begin{tabular}{cll}
                                                                                  & \multicolumn{1}{c}{\begin{tabular}[c]{@{}c@{}}Structured\\  Learning\end{tabular}}               & \multicolumn{1}{c}{Q-Learning}                                                                                                  \\ \cline{2-3} 
\multicolumn{1}{c|}{State Space}                                                  & \multicolumn{2}{c|}{Common$\colon \stsp$}                                                                                                                                                                                                              \\ \cline{2-3} 
\multicolumn{1}{c|}{Action Space}                                                 & \multicolumn{2}{c|}{Common$\colon \acsp$}                                                                                                                                                                                                              \\ \cline{2-3} 
\multicolumn{1}{c|}{Rewards}                                                      & \multicolumn{2}{c|}{Common$\colon R(s,a)$}                                                                                                                                                                                                              \\ \cline{2-3} 
\multicolumn{1}{c|}{STM}                                                          & \multicolumn{1}{c|}{\begin{tabular}[c]{@{}c@{}}Explicitly defined,\\ parameterized\end{tabular}} & \multicolumn{1}{c|}{Not needed}                                                                                                 \\ \cline{2-3} 
\multicolumn{1}{c|}{\begin{tabular}[c]{@{}c@{}}Policy \\ Generation\end{tabular}} & \multicolumn{1}{c|}{\begin{tabular}[c]{@{}c@{}}Solver invoked at \\ runtime\end{tabular}}        & \multicolumn{1}{c|}{\begin{tabular}[c]{@{}c@{}}Temporal difference \\ equation used to \\ estimate Q function\end{tabular}} \\ \cline{2-3} 
\end{tabular}
\label{tab:stlr_v_ql}
\end{table}

A summary of the differences between the controllers is shown in
Table~\ref{tab:stlr_v_ql}.  The MDP components that are common between the two
controllers are: the discrete state space $\stsp$, the discrete action space
$\acsp$, and the reward function $R(s,a)$. The other components of the
controllers are different, namely the STM and policy generation method. The
Structured Learning controller contains a parameterized state transition matrix
and employs an MDP solver at runtime to generate a control policy. In contrast,
the Q-Learning controller does not contain an explicit state transition matrix,
and instead contains a running estimate of the MDP's Q function. This allows
the Q-Learning controller to completely bypass having to store an STM and run
an MDP solver. 

One fundamental difference between the two techniques is that the Structured
Learning controller has an STM with a pre-populated structure, and only
parameters within the structure need to be learned at runtime. There are some
aspects of the STM that are fixed at design time, and in this way, some
structure does not need to be learned. In contrast, the Q-Learning controller
has no a priori structural assumptions, and must learn the entire Q function at
runtime.  Depending on how much of the STM is known at design time in a given
application, the Q-Learning controller may have to learn more parameters at
runtime compared to the Structured Learning controller. 

\begin{table}[h!]
\centering
\caption{Number of parameters to learn at runtime for each controller.}
\label{tab:num_learn_params}
\begin{tabular}{cc}
\hline\noalign{\smallskip}
 Structured & \\
 Learning & Q-Learning  \\ 
\noalign{\smallskip}\hline\noalign{\smallskip}
$0 \leq \mxcard{\Theta} \leq {({N_S}-1){N_S}{N_A}}$ & ${N_S}{N_A}$ \\
\noalign{\smallskip}\hline
\end{tabular}
\end{table}

This contrast is shown in Table~\ref{tab:num_learn_params}, where $\Theta$ is
the number of unknown or time varying parameters in the STM, as was defined in
Equation~\ref{eq:structured_learning_stm}. This motivates a general design
guideline to be considered: to determine what type of learning approach to use,
an engineer should first analyze the percentage of the STM that is known versus
the percentage that is unknown or expected to be time-varying.  In other words,
the designer should seek to minimize the number of parameters that need to be
learned at runtime, which is $\mxcard{\Theta}$ in Structured Learning, and
${N_S}{N_A}$ in Q-Learning  (recall that $\mxcard{\Sigma}$ denotes the
cardinality of a set $\Sigma$).  If many parameters out of the STM need to be
learned at runtime, it is possible that $\mxcard{\Theta} \geq {N_S}{N_A}$, in
which case the Q-Learning approach would have less parameters to learn at
runtime and likely provide better adaptation performance.

The two different controller designs investigated in this section were 
formulated to concretely explore trade-offs between model-based (Structured 
Learning) and model-free (Q-Learning) in a tangible, real-world example.
In the remainder of this Section we first detail the common components of the 
two techniques, and then elaborate on their differences. 

\subsection{State and Action Spaces} 

Both controllers utilize a multivariate state space, a concept 
introduced in Section~\ref{subsec:factorization}. The MDP state space is 
essentially the combination of each of the states of the sensor 
application, packet queue and LTE-M modem. This is defined as shown in 
Equation~\ref{eq:state_space}:

\begin{equation}
\label{eq:state_space} 
\begin{split}
s = (s_a, s_q, s_m) \in \stsp \\
s_a \in \{0,1,..,N_{SA}-1\},\\
s_q \in \{0,1,...,N_{SQ}-1\},\\
s_m \in \{0,1,..,N_{SM}-1\},
\end{split}
\end{equation}

\noindent where $s \in \stsp$ is the state variable, which is composed of there 
separate variables $s_a, s_q, s_m$: the sensor application state, queue state 
and modem state, respectively. 

In general, we are interested in 
applications that can run in one of a set of alternative modes 
(see Section~\ref{sec:application}). Thus we define $s_a$ 
as the application mode, a state variable taking a value out of a 
discrete set of $N_{SA}$ modes. 
The packet queue has a specific number of packets in it at any given 
time. In order to allow the controllers to make decisions based on the 
current state of the queue, we define $s_q$ as the number of packets in 
the queue and make it part of the MDP state space. 



We assume that the queue is of a fixed size $N_{SQ}-1$, due to our 
goal of housing it in a resource constrained MCU. The queue can only 
hold up to $N_{SQ}-1$ packets, and any attempts to queue additional packets 
beyond this limit will result in the newest packet being discarded. The
discarding of a packet represents a loss of data and is a very undesirable 
event, that our controllers must seek to avoid through the decisions in their 
respective control policies. 

Note that in other applications, some amount of data loss may be tolerable. 
Such applications can be accommodated readily in both controller designs by 
making suitable adaptations. We omit details of these adaptations in this paper 
for brevity.

The LTE-M modem is a complex mixed-signal System-On-Chip 
(SoC), containing the LTE-M protocol implementation and runtime signal 
processing. There is an enormous state space that could be defined for 
the inner workings of the modem. However, for our controllers the majority 
of that information is not relevant and thus we 
collapse the modem state into a small set of states that is detailed 
enough for the controller to implement a high performing control policy, 
without being overly burdened with the computational and storage 
implications of a large state space. In this spirit, we leverage the 
concept of Transition States introduced in 
Section~\ref{subsec:transition-states}, and collapse the large number of 
fine-grained LTE-M modem states into three coarse-grained 
states: ${M\_OFF}$, ${M\_CONNECTING}$, and ${M\_CONNECTED}$. 
The ${M\_CONNECTING}$ state is a transition state, and 
the other two are not. 

${M\_OFF}$ refers to the modem being fully powered off, and the modem 
can remain in this state indefinitely until commanded otherwise. 
${M\_CONNECTING}$ is the state that begins immediately after the modem 
has been powered on and ends when a successful Internet connection has 
been established. The modem cannot remain indefinitely in this state. By 
definition, it is guaranteed to transition out of this state after a 
specified amount of time steps. ${M\_CONNECTED}$ is the state when a 
working Internet connection has been established and continues to be 
maintained. Transmission of packets is only possible in the 
${M\_CONNECTED}$ state. This modeling approach forces us to pick a scalar 
constant for the modem's power consumption in each of the three states. 

\begin{sloppypar} 
As can be observed from Figure~\ref{fig:ltem-power}, the power 
is roughly constant in the ${M\_OFF}$ and ${M\_CONNECTED}$ states. 
However, this is not the case in the ${M\_CONNECTING}$ state. We address 
this by representing power consumption during the entire transition as a 
fixed value: the average power consumption during the duration of the 
transition. This simplification is a way of providing the MDP the 
information that it needs to implement a high performing policy, in as 
compact a representation as possible. This modeling approach is a design 
choice, and we note that an interesting area for future study is in the 
trade-offs for varying levels of modeling expressiveness in this area. 
\end{sloppypar} 

In this system, the controllers are being tasked with turning the LTE-M 
modem on and off. This binary control implies an ``on'' action that powers 
on the LTE-M modem and commands it via the modem's interface to attach 
to the cell tower and establish an Internet connection. Conversely the 
``off'' action implies tearing down any existing Internet connections, and 
shutting off the LTE-M modem gracefully via the modem's shutdown 
procedures. The resulting action space is shown in 
Equation~\ref{eq:action_space}. The two actions 
$\mathit{off}$ and $\mathit{on}$ in Equation~\ref{eq:action_space}
are abstractions of multi-step LTE-M modem command sequences. 

\begin{equation}
\label{eq:action_space} 
a \in \acsp = \{\mathit{off},\mathit{on}\}
\end{equation}

\subsection{Rewards}
In order to ``motivate'' the controllers to find an effective balance to the 
latency versus energy efficiency trade-off described in 
Section~\ref{sec:application}, we define the reward function as shown 
in Equation~\ref{eq:rewards}. The reward function maps each state-action 
pair $(s,a)$ to a scalar reward.

\begin{equation}
\label{eq:rewards}
\begin{split}
R(s,a) = r_1 I(s,a) + r_2 N_T(s,a) + r_3 N_D(s,a),\\
r_1=-10, \\ 
r_2\in \{3,4,...,10, 100, 1000\}, \\ 
r_3=-100,
\end{split}
\end{equation}

\noindent where $I(s,a)$ is the average electrical current consumed by 
the modem, $N_T(s,a)$ is the number of packets known to be transmitted, and 
$N_D(s,a)$ is the number of packets dropped by 
the modem due to an overflowing queue in the previous timestep. For each of 
these quantities we use the function arguments $(s,a)$ to denote the respective
value of each of the terms known, expected, or averaged when action $a$ is 
taken in state $s$. 
Instead of power consumption, we use electrical current in its place due to it 
being equally suitable (given a constant voltage) and more 
straightforward to measure with an MCU in an embedded system. 

With this formulation, the scalar reward is thus a linear combination of 
observable time-varying signals and quantities. This formulation steers 
both controllers to our desired goals, by rewarding them (with a 
positive reward value) when a packet is transmitted successfully and 
penalizing them (with a negative reward value) when electrical current 
is consumed or the packet queue overflows. The reward constants 
$r_1,r_3$ were selected via experimentation and $r_2$ was left as a free 
parameter in order to be able to generate a set of instances for each 
controller. Each instance in the set places different amounts of 
importance on the latency requirement relative to the energy efficiency 
requirement. This approach allowed us to simulate a suite of controllers 
for each method, and plot the resulting performance for a more robust 
comparison. The resulting policies are ones where the respective 
controllers turn the modem on and off at each discrete time step, in the 
way they determine is the optimal approach for obtaining the maximal 
rewards. In other words, they attempt to transmit the packets generated by the 
sensor application without incurring undesired delay or consuming more 
electrical power than is needed, through dynamic and changing conditions.

In Section~\ref{subsec:controller1} through Section~\ref{subsec:controller2}, 
we detail the differences between the two controllers that we are evaluating in 
this case study.

\subsection{Structured Learning Controller} 
\label{subsec:controller1} 

The Structured Learning controller consists of the common components 
described above, plus the addition of the STMs and an MDP solver. The 
STMs are described in this section, and the solver is described in 
Section~\ref{sec:implementation}. 

The stored STMs at any given time are a combination of constants and 
time-varying parameter estimates. The constants are programmed in at design
time, and the estimates are maintained by 
observing samples of the relevant quantities, and using the Temporal 
Difference (Equations~\ref{eq:td}) method to update the estimates. 
The estimates are plugged into
the STM data structures, which serve to maintain fully populated STMs
at each time step. 

The STMs are constructed using a factored formulation, 
which greatly reduces the storage requirements of the MDP. 
The factorization procedure is shown in 
Equation~\ref{eq:stm_factorization_1} through Equation~\ref{eq:stm_factorization_5}. 
The factorization serves to convert one large multivariate conditional 
probability distribution into a product of terms functions, each having the form of a
lower dimensional conditional probability distribution. The terms 
correspond to the subsystems of the sensor application, packet queue and 
LTE-M modem, respectively. This re-arrangement causes a significant 
reduction on the MDP storage requirements. 

\begin{equation}
\label{eq:stm_factorization_1} 
p(s^{(n+1)} | s^{(n)}, a^{(n)}) = p(s_{a}^{(n+1)},s_{q}^{(n+1)},s_{m}^{(n+1)}| s^{(n)}, a^{(n)})
\end{equation}

\begin{equation}
\label{eq:stm_factorization_2}
= p(s_{a}^{(n+1)} | s_{q}^{(n+1)},s_{m}^{(n+1)}, s^{(n)}, a^{(n)}) \cdot p( s_{q}^{(n+1)},s_{m}^{(n+1)} | s^{(n)}, a^{(n)})
\end{equation}

\begin{equation}
\label{eq:stm_factorization_3} 
= p(s_{a}^{(n+1)} | s_{a}^{(n)}) \cdot p( s_{q}^{(n+1)},s_{m}^{(n+1)} | s^{(n)}, a^{(n)})
\end{equation}

\begin{equation}
\label{eq:stm_factorization_4}
= p(s_{a}^{(n+1)} | s_{a}^{(n)}) \cdot p( s_{q}^{(n+1)} | s_{m}^{(n+1)}, s^{(n)}, a^{(n)}) \cdot p(s_{m}^{(n+1)} | s^{(n)}, a^{(n)})
\end{equation}

\begin{equation}
\label{eq:stm_factorization_5} 
 = p(s_{a}^{(n+1)}| s_{a}^{(n)})\cdot{}p(s_{q}^{(n+1)}| s^{(n)}, a^{(n)})\cdot{}p(s_{m}^{(n+1)}| s_{m}^{(n)}, a^{(n)})
\end{equation}

\subsubsection{Sensing Application} Given the observability of the 
sensing application's mode, the controller can maintain parameters that 
statistically characterize how often the application is in a given mode, and 
how likely the CPS is to transition from any mode to any other given mode. 
These characterization parameters are listed 
in Equation~\ref{eq:app_params}. Using the latest values of these 
parameters, the $s_a$ term of the factored STM can be fully 
instantiated. 

\begin{equation}
\label{eq:app_params} 
\begin{split}
\hat{\sigma}_{i,j} = p(s_{a}^{(n+1)}=j | s_{a}^{(n)}=i) \forall (i,j) \in \{0,1,...,N_{SA}-1\}^2 \\
\end{split}
\end{equation}

\subsubsection{Packet Queue}

Since it is part of the state space, the dynamics of the queue must also 
be modeled as a transition matrix. This results in having to model a 
fully deterministic process into stochastic structures in order to fit 
into the MDP framework, and thus most of the probabilities are either 1 
or 0. The transition probabilities are almost all known at design time, 
since the dynamics of the queue do not change. The only uncertainty 
comes from the rate of packets entering the queue from the sensing 
application, and the rate leaving the queue from the LTE-M modem 
connection. 

The packet queue's state is defined as the number of packets in 
it at a specific timestep. The transition probabilities amount 
to the likelihood of transition to another state, in other words the 
change in the number of packets. The transition to a state where more 
packets are inserted corresponds directly to the sensing application's 
packet generation rate. These events are combined with the probability 
of packets being removed from the queue by being transmitted to the 
cloud server. If the modem state and action are such that the modem is 
not yet connected, then no packets can leave the queue and thus 
transitions to states where the number of packets is reduced are not 
possible. If the modem is connected, then packets can leave the queue. 

\subsubsection{LTE-M Modem} 

\begin{figure} 
\centering 
\includegraphics[width=4in, angle=0]{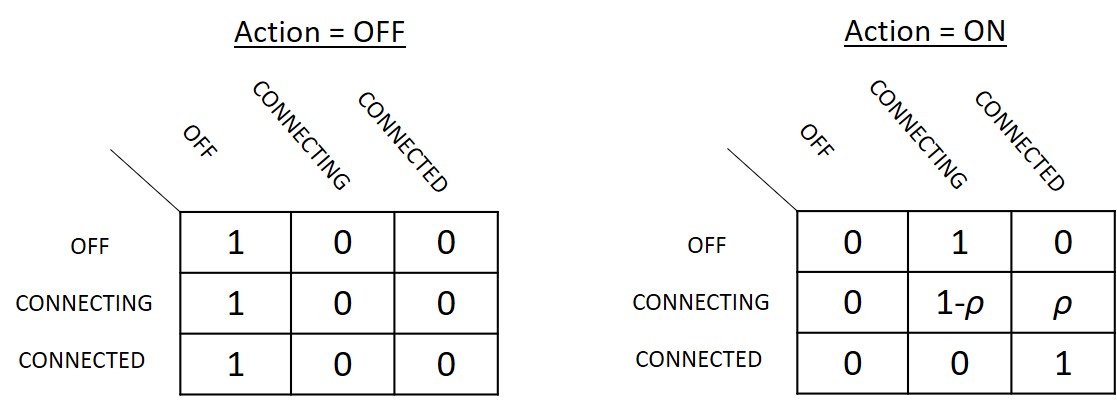} 
\caption{State transition matrices for LTE-M Modem. \\ 
 Only the parameter $\rho$ is learned at runtime. \\ The remaining entries 
 are specified at design time.}
\label{fig:modem_stm} 
\end{figure}

In our model, the dynamics of the LTE-M modem are a direct application of the 
Transition State concept. In order to instantiate this 
component of the STM, the controller needs to maintain a running 
estimate of how long the LTE-M modem takes to connect to the network. We 
refer to this time-varying quantity as $T_{C}$ and its most recent 
estimate as $\hat{T}_{C}$. With this estimate, the resulting transition 
probabilities are shown in Figure~\ref{fig:modem_stm}. Following the 
Transition States theory from the literature (e.g. ~\cite{sapi2018x1}), the 
value of the parameter 
$\rho$ is defined in Equation~\ref{eq:ts_param}, where $T_{F}$ is the 
duration of one control frame. 

As can be observed from the Figure~\ref{fig:modem_stm} and Equation~\ref{eq:ts_param}, 
this component of the STMs is 
parameterized by the running estimate $\hat{T}_{C}$ (through $\rho$) in two 
elements, and combined with known constants for the remaining elements.  

\begin{equation}
\label{eq:ts_param} 
\hat{\rho} = \left[\mathit{floor}\left(\frac{\hat{T}_{C}}{T_{F}}\right)\right]^{-1}
\end{equation}

\subsection{Q-Learning Controller} 
\label{subsec:controller2} 

The Q-Learning controller was implemented directly from the description 
of the technique in~\cite{sutt1998x1}. In this method, a function $Q$ is 
created as a mapping $Q(s,a) : (S \times A) \rightarrow \mathbb{R}$,
where $\mathbb{R}$ denotes the set of real numbers.
Each mapping in the function represents an estimate of the total amount 
of reward an agent can expect to accumulate over the future, starting 
from a given state $s$ and taking a given action $a$. The $Q$ function is 
updated on each iteration of the controller using the Temporal 
Difference. Using the latest 
estimated version of $Q(s,a)$, the action is selected by comparing all 
actions for the given state $s$ and selecting the action with the 
largest value. The remaining details of the Q-Learning method can be found
in~\cite{sutt1998x1}. 

\section{Simulation}
\label{sec:simulation}

\begin{figure} 
\centering 
\includegraphics[width=4in, angle=0]{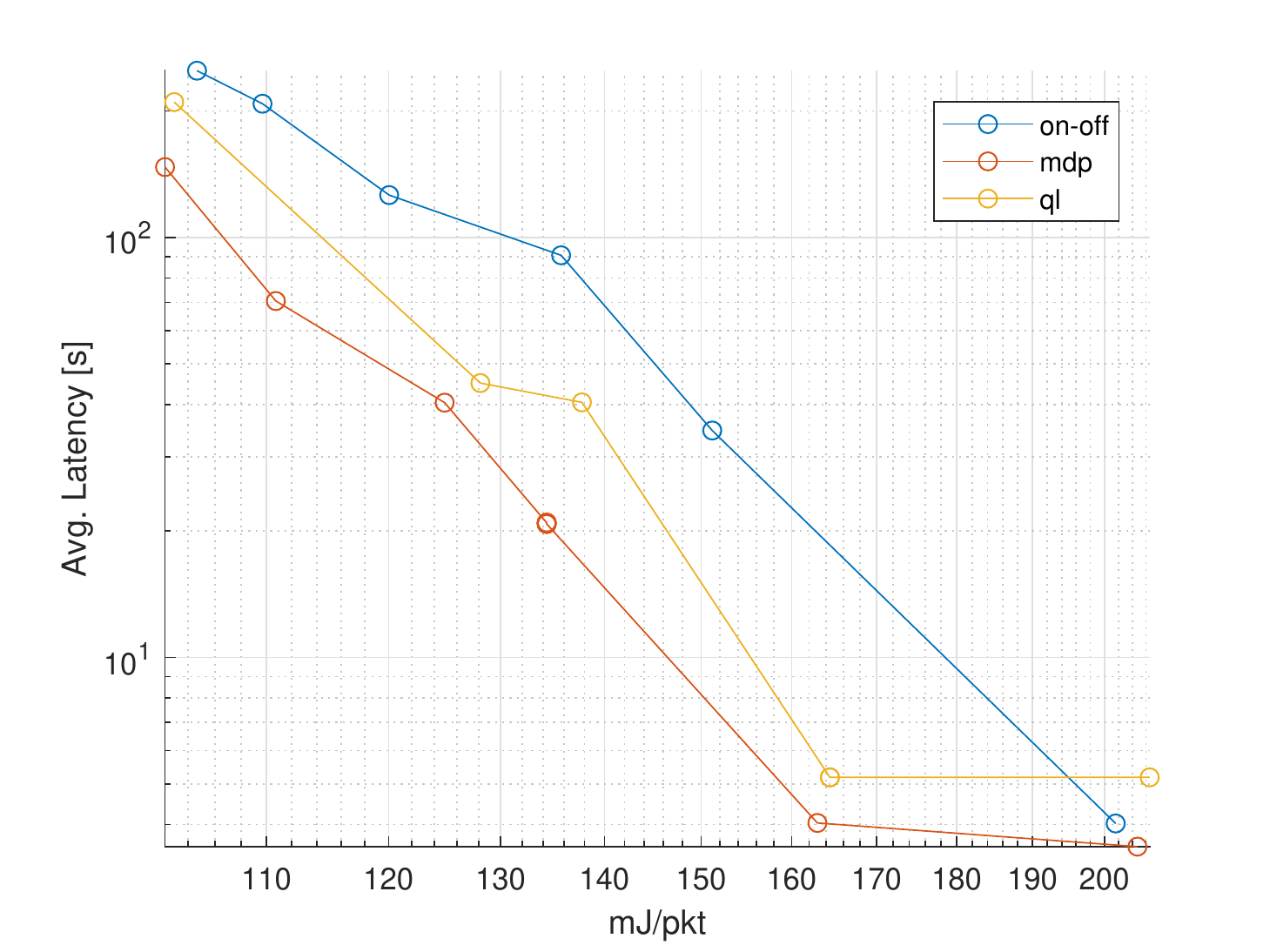} 
\caption{Simulation Results: Energy efficiency versus communication latency.}
\label{fig:sim_results} 
\end{figure}

In order to objectively compare the runtime performance of the 
Structured Learning and Q-Learning controllers presented in Section~\ref{sec:mdp},
we created a MATLAB simulation containing models of all the 
subsystems described in the case study. In our simulation testbed, a sensor application generates 
packets at rates consistent with a given mode, and also simulates the 
transition between modes at specified transition rates. A packet queue 
object models a generic fixed length queue, which acts as a First In 
First Out (FIFO) data structure, and overflows if the maximum number of 
elements is exceeded. A stochastic model of the LTE-M modem was created 
using the collected time-series data and electrical power measurements (see
Section~\ref{subsec:ltem}).

The simulation was run with three separate controllers: the two 
MDP-based controllers described in the previous section, as well as a 
third manually-generated policy. The manually-generated policy simply 
checks the queue, and turns on the modem any time a specified number $N_q$ of 
packets are in the queue, where $N_q$ is a parameter of the policy. 
Once the modem is turned on, it remains on until the queue is fully 
drained. 

The simulation was run multiple times for each controller, and each 
simulation run is represented by a single point in the graph of 
Figure~\ref{fig:sim_results}. The fixed threshold (manually generated), Structured Learning, 
and Q-Learning approaches are denoted by the ``on\-off'', ``mdp'', and ``ql'' 
traces, respectively. The best performance corresponds to points that 
have the lowest average communication latency (the vertical axis) and 
simultaneously the lowest average energy efficiency (the horizontal 
axis). 

For the fixed threshold technique, we generated multiple policies by 
varying the $N_q$ threshold at which the modem was powered on. For the 
MDP-based controllers, we varied the $r_2$ constant in the reward 
function from Equation~\ref{eq:rewards}. This approach gave us a set of 
control policies for each controller, allowing us to fully 
explore the performance limits of each technique. 

Our first conclusion from the data in Figure~\ref{fig:sim_results} is that both MDP-based 
controllers outperform the fixed threshold approach, for all 
possible values of the $N_q$ threshold. We believe this to be due to 
richer and more expressive policies generated by the MDPs; while the 
fixed threshold policies are a function of the packet queue state only 
(ignoring the application and modem characteristics), the MDP-based 
policies materialized as a function of the entire state space. In this case study, 
since the MDP is able to reason using algorithmic methods on data 
structures and computations on conditional probabilities, it can 
consider more effects and consequences systematically and produce highly 
optimized policies that are more expressive relative to the simple, 
manually-derived, fixed threshold heuristic.

Our second conclusion from the data is that the Structured Learning controller
outperforms the Q-Learning controller. As an example, if we tune the rewards
such that both learning controllers achieve an average packet transmission
latency of 20 seconds, the Structured Learning controller is able to accomplish
this with an average energy efficiency of 135 mJ per packet, compared to 163 mJ
per packet on the Q-Learning controller. This amounts to a 17\% savings in the
transmission energy for sending the exact same packets at the same average
latency. This can be an important difference since transmission energy is often
the largest source of energy consumption on an LTE-M connected sensor. 

We believe that Structured Learning outperforms Q-Learning in this example 
because we have focused its learning on only the time-varying aspects of 
the system (e.g. modem power, cell tower connection time, etc.) and 
have designed it to accept as unquestionable truth the other dynamics 
and attributes (e.g., that the packet queue contains one less packet 
after a packet is removed from it, etc.). In contrast, the Q-Learning 
controller is forced to learn (and continue to update indefinitely) all 
aspects of the system transition probabilities in response to selected 
actions. It must continually experiment with exploratory actions and 
accumulate data to learn all of the system dynamics 
(including well understood behavior, such as the packet queue's dynamics), 
and our results demonstrate that expending learning effort on such immutable 
aspects is both unnecessary and detrimental to the overall system performance.

\section{Implementation}
\label{sec:implementation}
In this section, we detail the results of implementation experiments 
performed to assess the viability of the competing MDP-based control 
strategies in the context of a state-of-the-art processing platform for 
resource-constrained CPSs. The alternatives are implemented on a typical MCU that would 
be used to realize our CPS case study, and are compared in terms of their 
execution time, memory usage and processing power consumption. 

\subsection{Experimental Setup} 

The competing controllers were implemented on the Silicon Labs EFM32GG, 
a small and low power ARM Cortex M3-based MCU. The processor was running 
on the EFM32 STK3700 development kit, which houses the CPU as well as 
sophisticated energy monitoring circuitry. The EFM32GG contains 128 kB 
of RAM and 1MB of FLASH, which make it a reasonably capable platform for the CPS in 
our case study at the time of this writing. 

In order to compare the controllers objectively, we created the 
following experimental setup on the EFM32GG development board. All 
controllers were implemented in C and stored in the MCU's program 
memory one at a time. Memory usage was computed by statically allocating 
all data structures and examining the map file that the MCU's compiler 
generates. A common test harness was written for the EFM32GG, which was 
driven by a periodic timer interrupt. The interrupt rate was configured 
to be 100ms, which we use as the fixed-period discrete-time iteration 
rate for the controllers. 

The C program initially puts the CPU into its low power sleep mode. It 
remains in that mode until the periodic interrupt fires. Once the 
interrupt fires, the MCU is woken from sleep and it then executes the 
computations needed for one iteration of the controller under test. Once 
the iteration has completed, the MCU returns back to sleep mode where it 
waits for the next firing of the periodic interrupt. Since the sleep 
current is extremely low compared to the run current (microAmps compared 
to milliAmps), this approach allowed us to precisely measure both the 
execution time and computational energy required to execute each 
controller on the MCU by observing the current versus time profile of 
each controller. Real-time MCU current consumption was measured by using 
the EFM32GG board's energy monitoring tools, which allow very accurate 
current versus time data to be observed in the form of a high resolution 
time-series waveform capture. 

Using this simple fixed rate scheduling scheme combined with the 
Cortex-M3 sleep modes and the development board's current monitoring 
tools, we were able to observe the execution time and processing current 
consumed by the CPU for each control policy. This testbench provided a 
highly repeatable experimental setup where all settings were kept the 
same from case to case with the only difference being the control policy 
being used. 

\subsection{Matrix Format} 

\begin{figure} 
\centering 
\includegraphics[width=4in, angle=0]{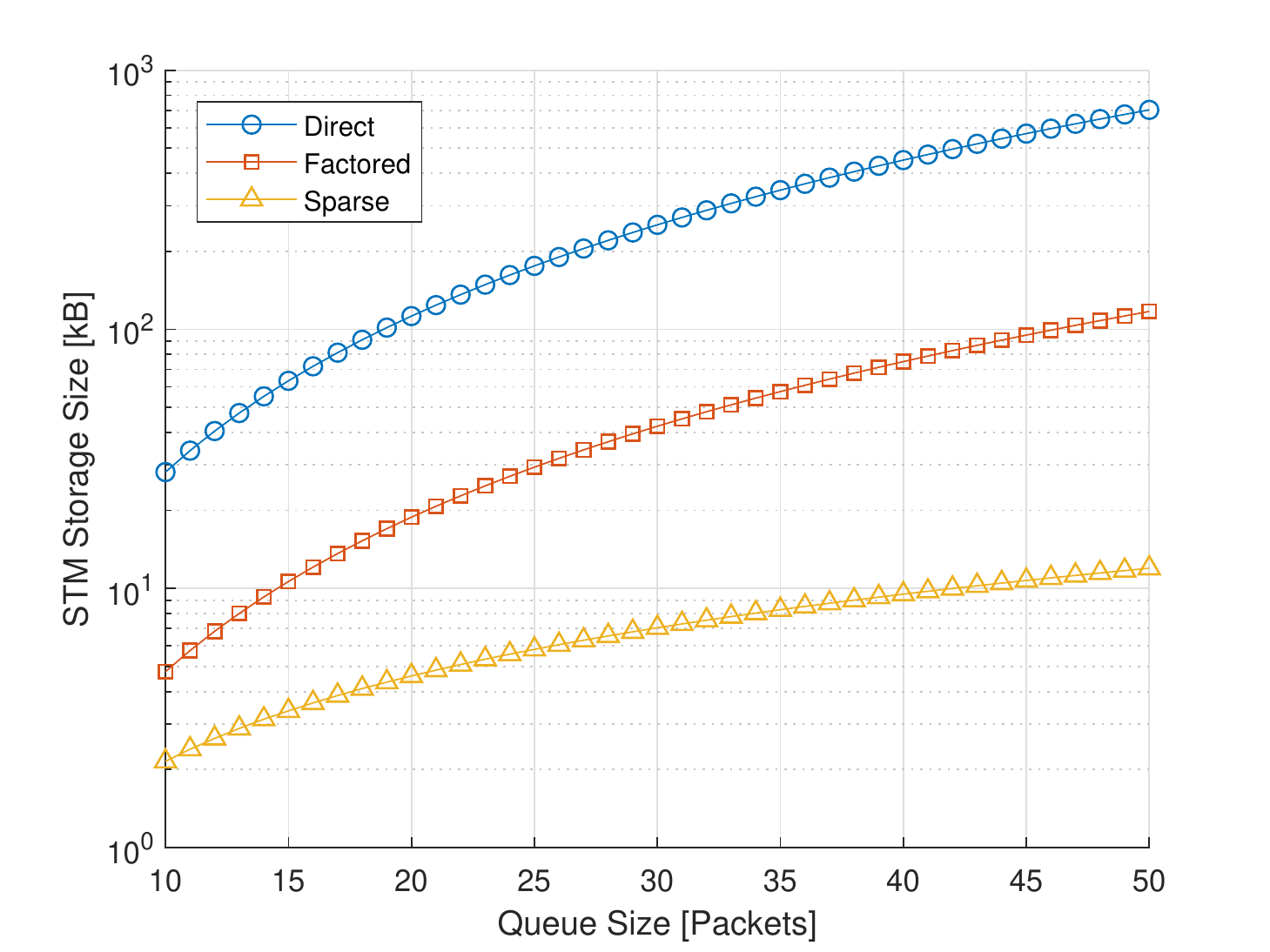} 
\caption{Transition matrix storage sizes.}
\label{fig:stm_size} 
\end{figure}

In our case study's MDP, the number of states $N_S$ is 66, and the number of
actions $N_A$ is 2. The number of non-zero elements $K_{NZ}$ in the STMs is
444.  This represents a sparsity of 94.9\%. Aside from the direct
implementation of the full STMs, we evaluated two CMM techniques: a factored
implementation and a sparse implementation. In Figure~\ref{fig:stm_size}, we
show the resulting STM storage sizes for these techniques on our case study,
over a range of packet queue sizes.

As can be observed from the data, both the factored and sparse implementations 
reduce storage size considerably. However, the sparse method is the most effective
in this regard and for this reason we selected this approach for the implementation. 

\subsection{Measurements} 

\subsubsection{Memory Usage}

\begin{table}[h!]
\centering
\caption{Data storage size in kiloBytes.}
\label{tab:storage_size}
\begin{tabular}{cccc}
\hline\noalign{\smallskip}
 & Structured  & Sparse Structured  &   \\
 & Learning (VI) & Learning (SVI) & Q-Learning  \\ 
\noalign{\smallskip}\hline\noalign{\smallskip}
\mxstm & 34.0 kB & 2.60 kB & - \\
Rewards & 0.51 kB & 0.51 kB & 0.51 kB \\
Q-Function & - & - & 528 B \\
\noalign{\smallskip}\hline
Total & 34.5 kB & 3.11 kB & 1.03 kB\\
\noalign{\smallskip}\hline
\end{tabular}
\end{table}

First, we compared the techniques in terms of how much data storage each 
required on the MCU. The results are shown in Table~\ref{tab:storage_size}, 
where the column labeled Sparse Structured 
Learning represents the Structured Learning method implemented with sparse 
matrices, as described in Section~\ref{subsec:svi}. 

We observe from Table~\ref{tab:storage_size} that the Q-Learning approach is the most 
favorable in this metric, and furthermore show that it requires 
significantly less data storage than the Structured Learning (VI) 
approach. However, when we apply the SVI method to Structured Learning, 
we see a huge reduction in the required data storage compared to VI. 
We conclude that in this case study, the CMM techniques reduce the 
storage requirements of Structured Learning to be much closer to that 
of Q-Learning, while not beating it in this regard. 

Generalizing these results beyond the case study, it can be seen from
Table~\ref{tab:storage_size} that all of the methods being compared require the
same amount of storage for the reward function. Thus, the difference in memory
usage is attributed to the storage of the STMs in the VI and SVI controllers,
compared to only the Q-Function in the Q-Learning controller. This difference
can be calculated in the general case as follows.  Assuming STM entries are 4
byte single precision floating point values, the Q-Function can be stored in
$4{N_S}{N_A}$ bytes, as it consists of a table of ${N_S}{N_A}$ floating point
values. The storage size of the STMs in the Structured Learning (VI) method is
$4{{N_S}^2}{N_A}$ bytes, as it consists of $N_A$ stochastic matrices, each of
size $({N_S}\times{N_S})$. 

The STMs in the Sparse Structured Learning (SVI) method were implemented with
sparse matrices stored in coordinate format~\cite{fili2017x1}. This format
stores only the non-zero elements of a matrix, along with two indices
representing the column and row index of the element, respectively.  Assuming
that $\lceil{\mathit{log}_2(N)/8}\rceil$ bytes are required to store an integer
index that can take on one of $N$ values, the storage size required for the
STMs in coordinate format is shown in Equation~\ref{eq:svi_stm_storage_size},
where $K_{\mathit{NZ}}$ is the number of non-zero elements in the STMs. 

\begin{equation}
\label{eq:svi_stm_storage_size} 
K_{\mathit{NZ}}(\lceil{\mathit{log}_2(NsNa)/8}\rceil + \lceil{\mathit{log}_2(Ns)/8}\rceil + 4)
\end{equation}

Evaluating the formula above (Equation~\ref{eq:svi_stm_storage_size}) using the
constants from the case study ($N_S=66$, $N_A=2$, $K_{\mathit{NZ}}=444$)
results in the storage sizes shown in Table~\ref{tab:storage_size}. The
formula can be used to predict the required storage sizes for other case
studies by appropriately changing the values of ($N_S,N_A,K_{\mathit{NZ}}$). 

\subsubsection{Computation}

\begin{table}[h!]
\centering
\caption{Execution time (in seconds), computation energy (in Joules) \\ and average power (in Watts).}
\label{tab:exec_time_power}
\begin{tabular}{cccc}
\hline\noalign{\smallskip}
 & Structured  & Sparse Structured  &   \\
 & Learning (VI) & Learning (SVI) & Q-Learning  \\ 
\noalign{\smallskip}\hline\noalign{\smallskip}
Per Control Iteration: & 3.5 $\mu$s / 117 nJ & 3.5 $\mu$s / 117 nJ & 211 $\mu$s / 7.06 $\mu$J  \\
Per Solver Iteration: & 50.1 s  / 1.67 J & 5.61 s / 187 mJ & - \\
\noalign{\smallskip}\hline
Average Power & 483 $\mu$W & 69.8 $\mu$W & 78.8 $\mu$W \\
\noalign{\smallskip}\hline
\end{tabular}
\end{table}

Next, we measured the execution time and power consumption of the MCU 
when executing the control algorithms for each of the competing techniques. We observe that 
Q-Learning requires the same computation on every control period. This 
consists of updating the Q-Function based on the observed state 
transition and reward, and computing the best action for a given state 
using the latest values in the Q-Function. The Q-Learning method 
performs these operations on every control period. 

In contrast, the 
Structured Learning techniques (VI and SVI) run a solver, which we invoked once per 
hour. These techniques compute a new control policy every hour, and the 
time and energy required to do this is shown in the second row of 
Table~\ref{tab:exec_time_power}. 
After computing a new policy every hour, the Structured Learning techniques simply look up which
action to use from a stored table for the remainder of that hour. 

From Table~\ref{tab:exec_time_power}, we see that the Structured Learning 
techniques (represented in the second and third columns of data) involve much 
less computation time and energy consumption during a typical control period 
relative to Q-Learning. Note that the first row of data for Structured Learning 
in Table~\ref{tab:exec_time_power} excludes the computation associated with 
solver execution, while the third row of data (labeled Average Power) includes 
the effects of solver computation.

We note that the Structured Learning implementation would likely require a 
priority-based pre-emptive scheduling scheme such that the control iteration
execution would take higher execution priority over the solver, such that any 
real-time deadlines associated with the controller are not missed due to
running the solver.

We conclude from the data in Table~\ref{tab:exec_time_power} that although 
Q-Learning does consume less average power (third row of data)
than Structured Learning (VI), when we apply the SVI method to 
Structured Learning we achieve less average power compared to 
Q-Learning. SVI reduces computation time by replacing standard matrix 
operations by sparse matrix operations. This results in a significant 
reduction in computation time, given that the STMs are extremely 
sparse. 

Generalizing these results beyond the case study, we identify the factors that
affect which of the methods is favorable in terms of computation costs ---
e.g., processing time and energy consumption.  In the Structured Learning
techniques, the size of the MDP and complexity of the STMs determine how long
it takes to execute a solver to produce a control policy. If the MDP is very
large and complex, the solver will take longer to execute. In contrast, the
Q-Learning technique is not affected at all by this attribute.  In this aspect,
it can be concluded that Q-Learning is better suited to deal with large MDPs
than Structured Learning in terms of computation expense. 

Another factor that is relevant is how often the Structured Learning techniques
are required to compute an updated policy. In general, a suitable update rate
is determined by the application's adaptation requirements, and how quickly the
time varying environment is changing.  In the case study presented here, the
policy is updated once per hour, but a system that adapts to more slowly
changing dynamics may only need to update the control policy once per day or
even less frequently. On the other hand, a system adapting to fast changing
dynamics may need to update the policy much more often, such as once per
second.  A faster update rate will generally increase the computational cost of
the Structured Learning techniques, whereas Q-Learning is not affected by this
consideration at all. In this regard, Structured Learning is better suited to
applications were the adaptation is on dynamics that are varying relatively
slowly in time. 

\begin{figure} 
\centering 
\includegraphics[width=5in, angle=0]{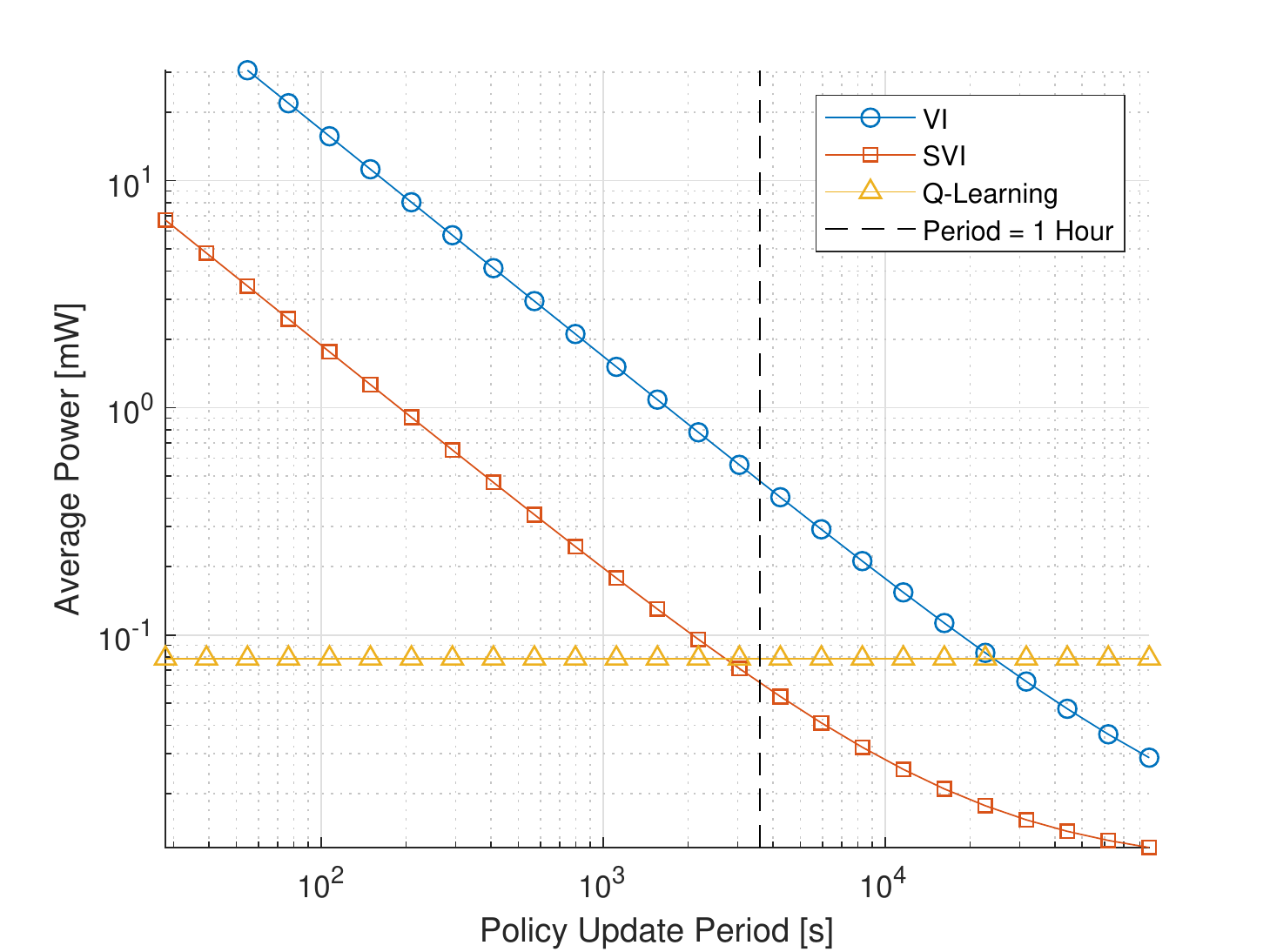} 
\caption{Average power consumed by the MCU on competing control algorithms \\ as a function of the policy update period.}
\label{fig:vary_epoch_period} 
\end{figure}

In this case study, the data point of a 1 hour update period leads to Sparse
Structured Learning having lower computational cost than Q-Learning.  The
crossover point where Q-Learning consumes less computational power is shown in
Figure~\ref{fig:vary_epoch_period} to be approximately at 45 minutes. It is
important to note that this crossover point is specific to the MDP size and
complexity (affecting solver execution time) and the choice of MCU (affecting
run current, sleep current, and solver execution time).

In summary, from the results of this and the preceding section, we see that 
Structured Learning with SVI is capable of applying control policies that are 
more robust compared to Q-Learning at a slightly lower average computational 
power consumption. Although the memory requirement overhead of SVI is 
greatly reduced compared to VI through the exploitation of sparsity, it is still 
slightly larger than that of Q-Learning.

\section{Conclusion}
\label{sec:conclusion}
In this work, we have provided a survey of recent developments in 
Compact MDP Models (CMMs), and by integrating several complementary CMM 
techniques, we have presented a novel CMM-based approach to CPS design. 
We have provided comparisons between CMM-based methods and Q-Learning in 
the context of CPS. The differences between the two approaches were 
explored conceptually, as well as through a detailed case study 
involving both simulation and a prototype implementation. 

From the results of this work, we conclude that Q-Learning can be 
considered a more robust technique when either very little is known 
about the system a priori, or a large percentage of the dynamics are 
expected to continually change at runtime. In contrast, when a 
significant portion of the system's environment or its dynamics are 
predictable, the CMM option can provide a substantially more efficient and 
robust approach. 

An LTE-M connected sensor was detailed as a CPS case study to compare a
CMM-based learning controller to an alternative controller that used
Q-Learning.  For a specified average packet transmission latency, the CMM-based
controller resulted in a 17\% reduction in LTE-M transmission energy, which is
often the largest source of energy consumption on an LTE-M connected sensor.
The energy savings are accomplished through strategic management of the LTE-M
modem and connection status, using learned dynamics of the system and its
environment. 

Since the learning controller must be implemented in the deployed system, and
its processing can be considered an overhead to the LTE-M connected sensor's
main purpose, we also analyzed the implementation costs of the two learning
controllers. The implementation was on a small microcontroller that is typical
of what would be used for such a CPS system. In this experiment, it was found
that the CMM-based controller used  69.8 $\mu$W compared to 78.8 $\mu$W for the
Q-Learning controller, an 11.4\% savings. However, the CMM-based controller
required more RAM to store its data structures --- 3.11kB compared to 1.03kB. 

Useful directions for future work include explorations into other 
challenging CPS case studies with larger state spaces, and continued 
development of compact techniques that provide self-awareness and 
runtime adaptation capabilities at all levels of embedded 
implementation.

\section{Acknowledgements}
\label{sec:ack}
This research was sponsored in part by the US National Science Foundation
(CNS1514425 and CNS151304).

\bibliographystyle{IEEEtran}
\bibliography{refs}

\end{document}